\documentclass[10pt, aps,prl,twocolumn,showpacs,superscriptaddress,nofootinbib,preprintnumbers,secnumarabic]{revtex4-2}

\usepackage{amsmath, graphicx}
\usepackage{xcolor}
\usepackage[colorlinks,linkcolor=blue,anchorcolor=blue,citecolor=blue,urlcolor=blue,]{hyperref}
\usepackage{orcidlink}
\usepackage{booktabs}
\usepackage{lineno}
\usepackage{float}
\usepackage{algorithm}
\usepackage{algpseudocode}
\usepackage{multirow}
\usepackage{diagbox}
\usepackage{array}
\usepackage{makecell}

\begin{document}

\title{Ultralight Bosons Explain the Mass–Spin Correlations in the Merging Binary Black Hole Population} 




\affiliation{School of Physics and Astronomy, University of Minnesota, Minneapolis, MN 55455, USA}
\affiliation{School of Physics, Anhui University, 111 Jiulong Road, Hefei, Anhui 230601, China }

\author{Xiao-Xiao Kou\,\orcidlink{0000-0002-7300-370X}}
\affiliation{School of Physics and Astronomy, University of Minnesota, Minneapolis, MN 55455, USA}

\author{Vuk Mandic\,\orcidlink{0000-0001-6333-8621}}
\affiliation{School of Physics and Astronomy, University of Minnesota, Minneapolis, MN 55455, USA}

\author{Ran Ding\,\orcidlink{0000-0002-2959-3140}}
\email[Contact Author:~]{dingran@mail.nankai.edu.cn}
\noaffiliation

\author{Chi Tian\,\orcidlink{0000-0002-5891-8573}}
\email[Contact Author:~]{ctian@ahu.edu.cn}
\affiliation{School of Physics, Anhui University, 111 Jiulong Road, Hefei, Anhui 230601, China }

\date{\today}

\begin{abstract}
Ultralight bosons could trigger superradiant instabilities in rapidly spinning black holes, forming oscillating clouds while extracting rotational energy.
We consider an extended, superradiance-informed spin distribution model that characterizes possible environment-induced spin variations and compare its predictions with the observed population of merging black hole binaries in the Gravitational-Wave Transient Catalogs (GWTCs). 
We find that the mass-spin relation predicted by a scalar boson with mass $m_b\sim 10^{-12} \,\rm eV$ is consistent with the GWTCs, with increasing significance from GWTC-3.0 to 5.0. The Bayes factor reaches $\ln \mathcal{B} \approx 7.8$ for GWTC-5.0.
Intriguingly, this mass range largely coincides with a previous study based on a waveform analysis of the GW190728 gravitational wave event. Our findings provide compelling evidence that a superradiance-informed spin distribution model is highly compatible with the expanding binary black hole population dataset.
\end{abstract}

\maketitle

\noindent\textbf{\textit{Introduction}} -- 
Gravitational wave (GW) astronomy has revolutionized the study of black hole (BH) astrophysics. Facilitated by the unprecedented sensitivity~\cite{KAGRA:2013rdx} of the LIGO-Virgo-KAGRA (LVK) observatories, the detection of hundreds of binary black hole (BBH) mergers has yielded a rapidly expanding Gravitational-Wave Transient Catalog (GWTC)~\cite{LIGOScientific:2018mvr,LIGOScientific:2020ibl,LIGOScientific:2021usb,KAGRA:2021vkt,LIGOScientific:2026tep,LIGOScientific:2026wfs} that serves as an invaluable probe for exploring fundamental physics governing BH formation and evolution~\cite{LIGOScientific:2020kqk,KAGRA:2021duu,Kimball:2020qyd,Zevin:2020gbd,Mapelli:2019bnp,Baibhav:2019gxm,Broekgaarden:2021efa,vanSon:2021zpk,LIGOScientific:2025pvj,LIGOScientific:2026ctl}. 
For example, the observed BBH population points to mass-dependent substructure in its spin distribution~\cite{LIGOScientific:2025pvj,Berti:2025usa,Tong:2025xir,LIGOScientific:2026ctl}, in particular a high-spin subpopulation at large masses~\cite{Wang:2022gnx,Pierra:2024fbl,Antonini:2024het,Sadiq:2025vly,Banagiri:2025dmy,Ray:2025xti,Sridhar:2025kvi,Bartos:2026xlt,Plunkett:2026pxt,Hussain:2026pfm,Alvarez-Lopez:2026ymo}, consistent with predictions for BBHs formed via hierarchical mergers in dense stellar environments and active galactic nuclei (AGN) disks~\cite{OLeary:2005vqo,Antonini:2016gqe,Rodriguez:2019huv,Doctor:2019ruh,Tagawa:2019osr,Mapelli:2020xeq,Tagawa:2020dxe,Tagawa:2021ofj,Gerosa:2021mno,Mahapatra:2024qsy,Li:2025iux}.
These findings present a challenge for accurately modeling such correlations within established frameworks of BBH formation and evolution.


Beyond astrophysics, these observations also open a unique, purely gravitational discovery window onto ultralight particles. Elusive ultralight bosons such as the QCD axion, axion-like particles, and dark photons are predicted in various beyond-the-Standard-Model scenarios and constitute compelling dark matter candidates~\cite{Arvanitaki:2009fg,Marsh:2015xka,Hui:2016ltb,Fabbrichesi:2020wbt}. When scattered off a spinning BH, ultralight bosons with wave frequencies satisfying the superradiance condition are amplified~\cite{Penrose:1971uk,Starobinskii:1973vzb}, extracting the BH's rotational energy through a Penrose-like process~\cite{Penrose:1971uk} and forming an exponentially growing macroscopic boson cloud around the BH~\cite{Press:1972zz,Bekenstein:1973mi,1980PhRvD..22.2323D}. The cloud can extract up to $\sim10\%$ of the BH's mass~\cite{East:2017ovw,Herdeiro:2017phl} and subsequently dissipates, as its energy is radiated via GW emission driven by either boson annihilation into gravitons or transitions between energy levels~\cite{Arvanitaki:2010sy,Yoshino:2013ofa}. These processes generate continuous quasi-monochromatic GW emission and leave imprints on BH spin distributions. The masses of ultralight bosons have been constrained by analyses of high-spin BBH events~\cite{Aswathi:2025nxa,Caputo:2025oap,LIGOScientific:2025brd,LIGOScientific:2026pwx}, BBH population spin distributions~\cite{Ng:2019jsx,Ng:2020ruv,Fernandez:2019qbj,Cheng:2022jsw,Ning:2026ebu}, continuous GW signals~\cite{Isi:2018pzk,Palomba:2019vxe,Zhu:2020tht,Mirasola:2025car,LIGOScientific:2025csr}, and stochastic GW backgrounds~\cite{Brito:2017zvb,Tsukada:2018mbp,Tsukada:2020lgt,Yuan:2022bem}. Notably, a recent study~\cite{Roy:2025qaa} reported tentative evidence for the existence of an ultralight scalar with mass $m_b\sim 10^{-12} \,\rm eV$ through waveform analysis of GW190728.

In this Letter, we show that ultralight scalar bosons could account for the mass-spin correlations observed in the BBH population by spinning down BHs through superradiance. Rather than assuming that all high-spin BHs are superradiantly spun down to fixed lower spins~\cite{Ng:2019jsx,Ng:2020ruv,Fernandez:2019qbj,Cheng:2022jsw,Ning:2026ebu}, we adopt an extended superradiance-informed spin model (hereafter, the \textit{superradiance spin model}). This model encodes the mass–spin correlations induced by superradiance while parameterizing uncertainties in the spin distribution arising from aspects of the BH environment other than the boson cloud. We use GWTCs to jointly infer the BBH mass function, redshift evolution, and spin distribution alongside the superradiance parameters, with the observation selection function evaluated self-consistently. 
Moreover, we treat the superradiance-active timescale, $T_{\rm age}$, as an event-specific source parameter, and we develop an appropriate method to marginalize over it. 

By performing hierarchical Bayesian inference~\cite{Thrane:2018qnx,Mandel:2018mve,Vitale:2020aaz} on LVK’s GWTC-3.0 to 5.0 catalogs, we find that the superradiance spin model consistently outperforms the Gaussian component spin model.
For GWTC-5.0, a scalar boson mass of $m_b=7.1_{-1.1}^{+1.8}\times10^{-13}\,\rm eV$ yields a Bayes factor of $\ln \mathcal B \approx 7.8$ relative to the Gaussian component spin model. 
Our results reveal the internal consistency between BBH population data and the BH superradiance, motivating future searches for signals of ultralight scalar bosons in this mass range.  Unless otherwise stated, we use units $G = \hbar = c = 1$ throughout this work.

\noindent\textbf{\textit{Baseline model}} --
Our analysis uses hierarchical Bayesian inference 
to constrain the population hyperparameters $\Lambda$ from catalogs of BBHs.
This method factorizes a population model $\pi(\theta\,|\, \Lambda)$ into independent mass, redshift, and spin sectors,  where $\theta$ denotes the single-event parameters (e.g., masses and spins). For the mass and redshift sectors, we adopt widely used parametric models: the primary mass and mass ratio follow the \textit{Broken Power Law + 2 Peaks} model, capturing a broken power-law continuum with overdensities near $10 M_{\odot}$ and $35 M_{\odot}$~\cite{LIGOScientific:2025pvj,LIGOScientific:2026ctl}, while the merger rate evolution follows the standard \textit{Power Law Redshift} model $\mathcal{R}(z) \propto (1+z)^{\kappa}$. Since superradiance affects only spin magnitudes and leaves the spin–orbit tilt distribution unaffected, we inherit the isotropic tilt-angle distribution, and modify only the spin-magnitude distribution. The baseline natal spin-magnitude distribution is the \textit{Gaussian component spin model}, which consists of a truncated normal distribution $\mathcal{N}_{[0,1]}(\mu_{\chi}, \sigma_{\chi})$ shared by both BHs. We then extend this baseline model to incorporate sthe uperradiant spindown effect as detailed in the next section.

\begin{figure}
    \centering
    \includegraphics[width=1.\linewidth]{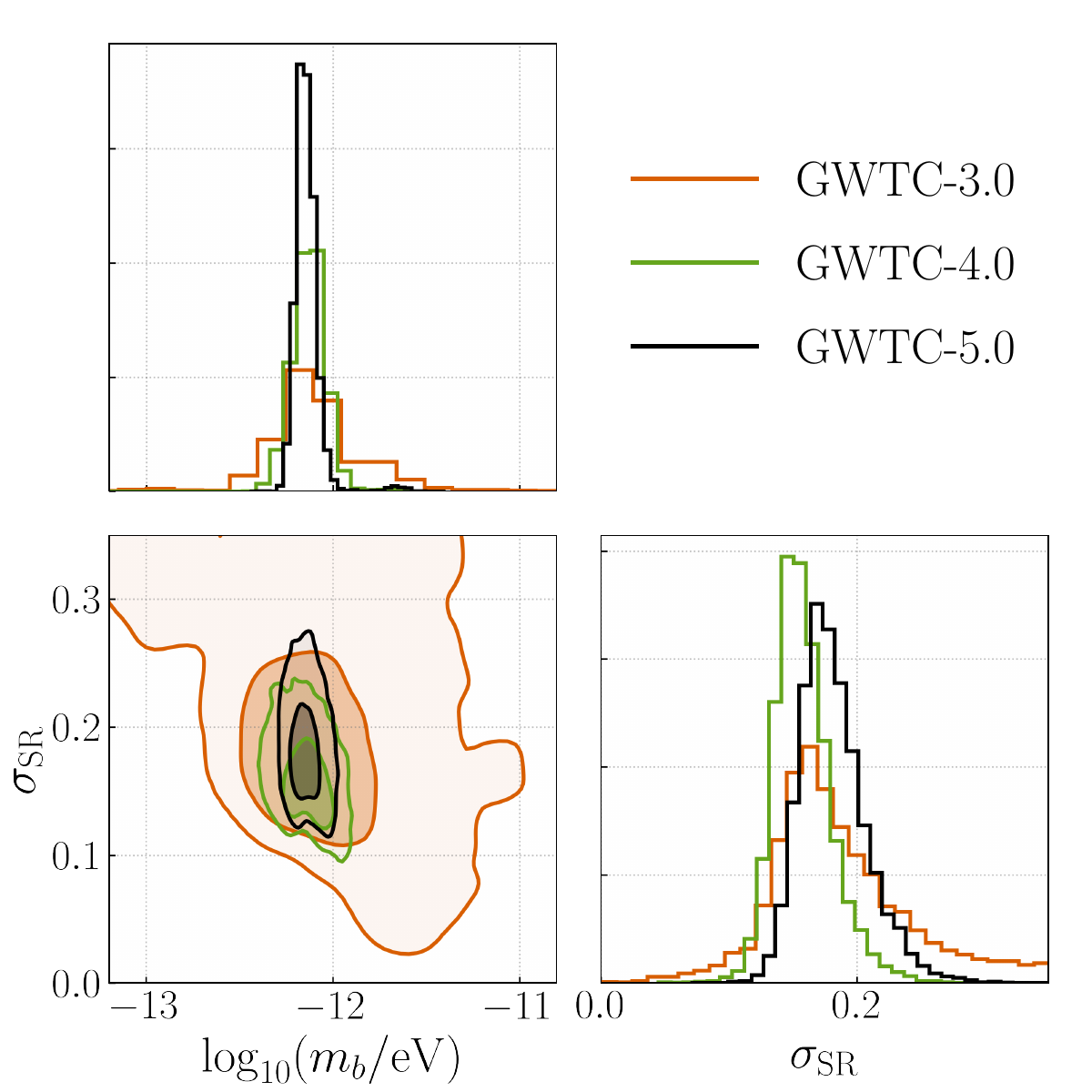}
    \caption{Posterior distributions for the boson mass $m_b$ and the environment-induced spin dispersion $\sigma_{\rm SR}$
    under the \textit{superradiance spin model} are shown for the GWTC-3.0 to 5.0. 
    Contours enclose $68\%$ and $95\%$ credible regions. 
    }
    \label{fig:fig1}
\end{figure}

\noindent\textbf{\textit{Superradiance spin model}} --
Bosonic fields of mass $m_b$ can form gravitationally bound states around the BH. These bound states are characterized by a complex frequency $\omega = \omega_R + i\omega_I$ ($\omega_I>0$), and grow exponentially in time when the superradiant condition $0 < \omega_R < m\,\Omega_H$ is satisfied. Here, $m$ is the azimuthal quantum number, $\Omega_H = \chi / [2M_{\rm BH}(1 + \sqrt{1 - \chi^2})]$ is the angular velocity of the outer BH horizon, and $\chi = J_{\rm BH}/M_{\rm BH}^2$ is the dimensionless spin parameter, where $M_{\rm BH}$ and $J_{\rm BH}$ denote the BH mass and angular momentum, respectively.

During the instability phase, the BH spins down, transferring mass and angular momentum to the boson cloud until the saturation condition $\omega \simeq m\Omega_H$ is satisfied, at which point the growth ceases. Over a timescale $T_{\rm age}$ between BH formation and binary merger, during which superradiance can operate, the existence of a fundamental ultralight boson $m_b \sim \omega_R$ imposes a mass-dependent maximum spin on isolated BHs, thereby defining a characteristic trajectory in the spin-mass plane. A superradiance-informed spin distribution model can be constructed by imposing this mass-dependent spin upper bound on a baseline spin distribution, depleting the region above $\chi_{\rm SR}$ where BHs would have been spun down. For a given final mass $M_f$, any BH with a spin exceeding the maximum  value $\chi_{\rm SR}(M_f, T_{\rm age}, m_b)$ is reassigned to this maximum spin.
As a result, a large number of BHs born with high spins accumulate near $\chi_{\rm SR}$, producing a sharp and narrow peak in the spin distribution~\cite{Fernandez:2019qbj}.

On the other hand, BHs are subject to various environmental effects that can further alter their spins. Extensive studies have investigated the combined effects of gas accretion and superradiance on supermassive BHs~\cite{Brito:2014wla,Hui:2022sri,Sarmah:2024nst,Nandakumar:2025aex,Giannakoudi:2026rks}. For stellar-mass BHs, accretion may occur through Roche-lobe overflow or stellar-wind capture from a binary companion~\cite{King:1999aq,Podsiadlowski:2002ww,Fragos:2014cva,Koninckx:2025kwy}, capture of material within an AGN disk~\cite{Tagawa:2020dxe,Chen:2023lxk,Mckernan:2017ssq,Kaaz:2021xla}, tidal disruption of nearby stars in dense stellar environments~\cite{Lopez:2018nkj}, or physical collisions with main-sequence stars in dense clusters~\cite{Rizzuto:2021atw,Kiroglu:2024xpc}. 
For example, stellar-mass BBHs embedded in AGN disks can accrete at or above the Eddington limit~\cite{Stone:2016wzz, Kaaz:2021xla, Bartos:2025pkv}, spinning up on a relatively short timescale of $\sim 10^7\,\rm yr$, comparable to the characteristic timescale of superradiance. These processes could introduce intrinsic scatter into the BH spin distribution, both during and after superradiance-driven spindown, causing BH spins to cluster around, rather than exactly at, $\chi_{\rm SR}$. 

To incorporate this effect, we adopt an extended spin distribution model motivated by BH superradiance, in which the sharp accumulation at $\chi_{\rm SR}$ is broadened to a truncated normal distribution of width $\sigma_{\rm SR}$, treated as a hyperparameter inferred from the data,
\begin{align}
\label{eq:spin-model}
&p(\chi\,|\,M_f, T_{\rm age}, m_b, \sigma_{\rm SR}, \Lambda_s) \\
&\qquad= \pi_{\rm base}(\chi\,|\,\Lambda_s)\,\Theta(\chi_{\rm SR}-\chi) + C\,\mathcal{N}_{[0,1]}(\chi_{\rm SR}, \sigma_{\rm SR})\,,\nonumber
\end{align}
where $\pi_{\rm base}(\chi\,|\,\Lambda_s)$ represents the baseline spin distribution,
$\Theta(x)$ is the Heaviside step function, and the spin hyperparameters $\Lambda_s$ include $\mu_{\chi}$ and $\sigma_{\chi}$ when the baseline model is the \textit{Gaussian component spin model}. The constant $C$ normalizes the total distribution to unity.\footnote{
Generally, a spin-up is accompanied by substantial mass accretion, which dynamically modifies the superradiant spindown rate. However, we expect that this effect has also been accounted for by $\sigma_{\rm SR}$.}.

The superradiant instability is most efficient when the dimensionless gravitational fine structure constant $\alpha_g$ satisfies $\alpha_g \equiv m_bM_{\rm BH} \sim \mathcal{O}(0.1)$. According to this criterion, ultralight bosons with masses in the range $m_b \sim 10^{-13.2}\,\mathrm{eV}$--$10^{-11}\,\mathrm{eV}$ influence the spins of BHs with masses between $\mathcal{O}(1\,M_\odot)$ and $\mathcal{O}(200\,M_\odot)$. In our analysis, we adopt a broader log-uniform prior on the boson mass, $\log_{10}(m_b/\mathrm{eV}) \sim \mathcal{U}(-14.5, -10)$, to encompass a wider range of $\alpha_g$, along with a uniform prior $\sigma_{\rm SR} \sim \mathcal{U}(0, 0.35)$. The priors for the remaining hyperparameters are identical to those adopted in Ref.~\cite{LIGOScientific:2025pvj,LIGOScientific:2026ctl}.

The duration $T_{\rm age}$ plays a central role in our analysis: superradiant spin-down is efficient only when $T_{\rm age}$ exceeds the characteristic timescale of superradiant instability. Since BBH formation can proceed through multiple channels whose formation-to-merger delay times plausibly span several orders of magnitude~\cite{Dominik:2013tma,Belczynski:2016obo,Rodriguez:2016kxx,Neijssel:2019irh,Bartos:2016dgn,Yang:2019okq}, a more realistic treatment allows $T_{\rm age}$ to take an independent value for each event in the catalog rather than imposing a common value across the population~\cite{Ng:2019jsx,Fernandez:2019qbj,Ng:2020ruv,Cheng:2022jsw}. We therefore promote $T_{\rm age}$ to an event-specific source parameter and marginalize over it, adopting a log-uniform prior $\pi(T_{\rm age})$ between $1\,\rm Myr$ and $2.2\,\rm Gyr$. The lower bound of $1\,\rm Myr$ is chosen to be shorter than the delay-time floor of all plausible BBH formation channels: isolated binary evolution typically yields $\gtrsim 10\,\rm Myr$~\cite{Dominik:2013tma,Belczynski:2016obo,Neijssel:2019irh}, and dynamical assembly channels give comparable or longer timescales\footnote{ Although the most rapidly merging systems can reach $\sim 10^5\,\rm yr$, they only represent an extreme tail of the distribution~\cite{Bartos:2016dgn,Yang:2019okq}.}~\cite{Rodriguez:2016kxx}. The upper bound of $2.2\,\rm Gyr$ is set by the cosmological consistency requirement that $T_{\rm age}$ plus the lookback time to the highest-redshift observed BBH event not exceed the age of the Universe. Our marginalization scheme for $T_{\rm age}$ is equivalent to drawing independent values from its prior and appending them to the existing event posteriors. As detailed in the Supplemental Material, this marginalization scheme is necessary to extract unbiased estimations of $m_b$.

Our dataset consists of 69 BBHs from GWTC-3.0~\cite{KAGRA:2021duu}, 153 BBHs from GWTC-4.0~\cite{LIGOScientific:2026tep} and 259 events from GWTC-5.0~\cite{LIGOScientific:2026ctl} with false alarm rates $\leq 1\,\mathrm{yr}^{-1}$. The Bayesian inference is carried out using \texttt{Dynesty}~\cite{Speagle:2019ivv} and \texttt{Bilby}~\cite{Ashton:2018jfp} together with the population likelihood~\cite{Mandel:2018mve,Vitale:2020aaz} as implemented in the open-source package \texttt{GWPopulation}~\cite{Talbot:2024yqw}. We use posterior samples from individual events consistent with the choice of Ref.~\cite{LIGOScientific:2026ctl} (see Supplemental Material for the waveform model choices). The survey sensitivity is estimated from the publicly available set of BBH injections recovered by the search pipelines across all four observing runs~\cite{Essick:2025zed}. Furthermore, we monitor the Monte Carlo uncertainty of the population likelihood~\cite{Essick:2022ojx}, and apply a smoothly-tapered likelihood cutoff in regions with insufficient Monte Carlo
support~\cite{Callister:2022qwb, Callister:2023tgi, Antonini:2024het,Antonini:2025ilj,Antonini:2025zzw, LIGOScientific:2026ctl,
Flanagan:2026ayy} (see Supplemental Material). 
We compute the maximum spin $\chi_{\rm SR}(M_f,T_{\rm age},m_b)$ using the public package~\texttt{SuperRad}~\cite{Siemonsen:2022yyf,May:2024npn}. Superradiance proceeds through the successive growth and saturation of unstable
azimuthal modes, and $\chi_{\rm SR}$ is the single spin value remaining after $T_{\rm age}$. This code employs fully relativistic templates to compute superradiant spindown for modes with azimuthal numbers $m\leq2$ and semi-analytical formulae to evaluate modes with azimuthal numbers up to $m\leq15$.

\noindent\textbf{\textit{Results}} --
We present the inferred scalar boson mass $m_b$ and the parameter $\sigma_{\rm SR}$ that characterizes the environment-induced variation, in Fig.~\ref{fig:fig1}, where clear peaks in the posterior distribution of $m_b$ can be identified. The posterior distributions of the boson mass are also significantly narrower for GWTC-4.0 and 5.0 than for GWTC-3.0. The $90\%$ credible intervals of $m_b$ and the corresponding Bayes factors for all three catalogs are reported in Table~\ref{tab:t1}. 
The higher Bayes factors obtained for GWTC-4.0 and GWTC-5.0 relative to GWTC-3.0 indicate stronger support for the mass-spin correlation predicted by the \textit{superradiance spin model} from these larger catalogs. In addition, as shown in Table~\ref{tab:t1}, we find that both the Bayes factors and $\sigma_{\rm SR}$ are marginally sensitive to two particularly loud events in GWTC-5.0, GW241011~\cite{LIGOScientific:2025brd} and GW250114~\cite{LIGOScientific:2025rid,LIGOScientific:2025wao,Chandra:2025jfc}. These events have exceptionally high SNRs and exhibit significantly high and low spins, respectively, placing them away from the $\chi_{\rm SR}$ trajectory (Fig.~\ref{fig:figM1}). 
After excluding these two events, $m_b$ remains unchanged, but the posterior distribution of $\sigma_{\rm SR}$ for GWTC-5.0 shifts toward a smaller value, while the Bayes factor further increases to $\ln \mathcal{B}\approx9.5$, exceeding that obtained for GWTC-4.0 (see the End Matter for further details). Other high-spin events, such as GW241113~\cite{LIGOScientific:2026pwx}, do not significantly affect our results. The parameter estimates for the baseline spin distribution, $\mu_{\chi}$ and $\sigma_{\chi}$, are presented in the Supplemental Material.

\begin{figure*}[ht]
    \centering
    \includegraphics[width=1.0\linewidth]{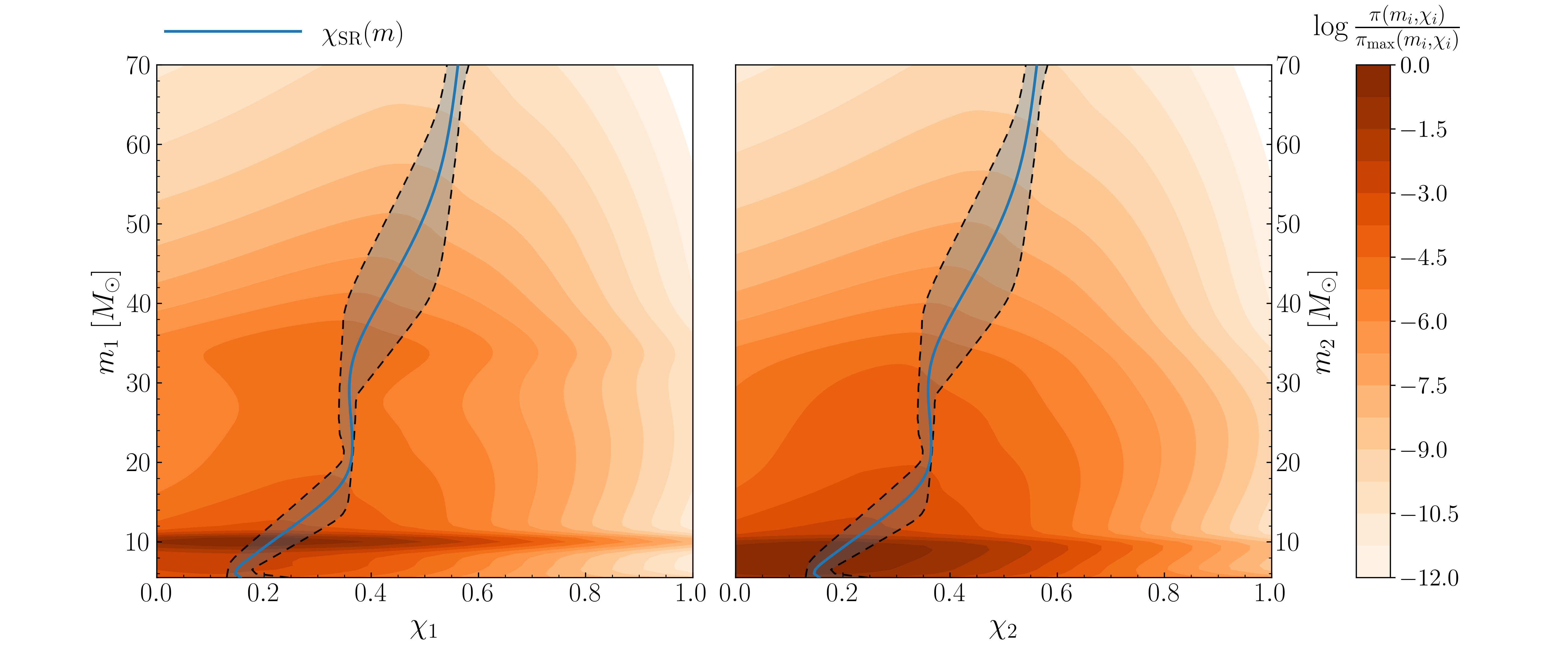}
    \caption{Reconstructed joint mass-spin distributions $\pi(m_1, \chi_1)$ (left) and $\pi(m_2, \chi_2)$ (right) for a scalar boson under the \textit{superradiance spin model}, inferred from the full GWTC-5.0 catalog and normalized to their maximum. Blue solid curves show the superradiance trajectory $\chi_{\rm SR}(m)$ at the posterior median for each case; dashed contours bracket the $90\%$ credible band from joint marginalization over $m_b$ and $T_{\rm age}$.
    }
    \label{fig:mass-spin}
\end{figure*}

\begin{table}[t!]
\centering
\renewcommand{\arraystretch}{1.22}
\begin{tabular}{c|c|c|c|c}
\hline\hline
Catalogs & Excluded events & $m_b\, (10^{-12}\,\rm eV)$ & $\sigma_{\rm SR}$ & $\ln\mathcal{B}$ \\
\hline
GWTC-3.0 & None & $0.74^{+1.75}_{-0.43}$ & $0.18^{+0.12}_{-0.07}$ &$2.8$ \\
\hline
GWTC-4.0 & None & $0.74^{+0.28}_{-0.22}$ & $0.16^{+0.04}_{-0.03}$ & $9.1$ \\
\hline
\multirow[c]{5}{*}{GWTC-5.0} 
& None & $0.71^{+0.18}_{-0.11}$ & $0.18^{+0.05}_{-0.03}$ &$7.8$ \\
\cline{2-5}
& \makecell{GW241011}& $0.69^{+0.22}_{-0.11}$ & $0.18^{+0.05}_{-0.03}$ & $8.1$ \\
\cline{2-5}
& \makecell{GW250114} & $0.72^{+0.16}_{-0.13}$ & $0.17^{+0.05}_{-0.03}$ & $8.4$ \\
\cline{2-5}
& \makecell{GW241011 \\ GW250114} & $0.71^{+0.15}_{-0.11}$ & $0.16^{+0.04}_{-0.02}$ & $9.5$ \\
\hline\hline
\end{tabular}
\caption{Inferred $90\%$ credible intervals of boson mass $m_b$, environment-induced spin dispersion $\sigma_{\rm SR}$ and Bayes factor against the baseline model for GWTC-3.0 to GWTC-5.0. 
}
\label{tab:t1}
\end{table}

According to robustness tests on mock data (see Supplemental Material), inference of the primary parameters $m_b$ and $\sigma_{\rm SR}$ is stable against internal variations introduced by marginalizing over $T_{\rm age}$'s, different catalog realizations, and the peak location of the BH mass function. In contrast, the baseline spin distribution hyperparameters, $\mu_{\chi}$ and $\sigma_{\chi}$ are less robustly constrained: their posteriors exhibit realization dependent variations. These hyperparameter distributions should therefore be interpreted with caution, whereas our reconstruction of $m_b$ can be regarded as reliable.


Fig.~\ref{fig:mass-spin} shows the reconstructed joint mass-spin distributions $\pi(m_1, \chi_1)$ and $\pi(m_2, \chi_2)$ for the primary and secondary BHs, respectively, under the \textit{superradiance spin model}, inferred from the full GWTC-5.0 catalog. The distributions are obtained by marginalizing the population model $\pi(m_i,\chi_i\,|\, \Lambda, T_{\rm age})$ over the hyperparameter posteriors and the prior $\pi(T_{\rm age})$. The $\chi_{\rm SR}(m_i)$ curves in Fig.~\ref{fig:mass-spin} and their uncertainty bands are computed  by evaluating $\chi_{\rm SR}(m_i, m_b, T_{\rm age})$ over posterior samples of $m_b$ and the prior on $T_{\rm age}$. In both panels, the reconstructed density is concentrated along the $\chi_{\rm SR}(m_i)$ curve, consistent with the model prediction. The pile-up is particularly pronounced near $m \sim 10\, M_{\odot}$, where the catalog provides the strongest constraints on the BH mass function~\cite{LIGOScientific:2025pvj, LIGOScientific:2026ctl}.

\noindent\textbf{\textit{Discussions}} --
In this Letter, we have shown that the mass-dependent spin distribution induced by ultralight scalar superradiance is a viable population model for the current BBH population data from GWTC-3.0 to GWTC-5.0. Adopting an extended superradiance-informed spin distribution that accounts for the environment-induced variations of BH spins, we recover a scalar boson mass $m_b\sim 10^{-12}\;\rm eV$ with a high Bayes factor relative to 
the baseline \textit{Gaussian Component Spins} model.

Appreciable exclusion regions in the scalar boson mass have been reported in previous studies of BBH populations~\cite{Ng:2019jsx,Fernandez:2019qbj,Cheng:2022jsw,Ning:2026ebu}, with only very weak support for a scalar boson with $m_b\sim 10^{-12}\,\rm eV$. These results were obtained by fixing the BBH mass distribution, neglecting spin selection effects, and relying on a delta-function spin model that
discards the mass-spin correlation within individual events. 
In contrast, by performing a comprehensive analysis that accounts for environmental effects and selection functions, while marginalizing over $T_{\rm age}$, we provide 
compelling evidence that the observed mass-spin correlations in the BBH population data can be explained by a superradiance-inspired spin model.

Notably, our findings are not in conflict with constraints from isolated high-spin BBH events~\cite{Aswathi:2025nxa,Caputo:2025oap,LIGOScientific:2025brd,LIGOScientific:2026pwx}, as our model allows a certain degree of variations in BH spins. Fixing $\sigma_{\rm{SR}}$ to 0.1 and reanalyzing the full GWTC-5.0 catalog, we find that $\ln\mathcal{B}$ decreases by an additional $\sim 2$, suggesting the importance of including this variation.
Using the full GWTC-5.0 catalog and a superradiance spin model that neglects the environmental spin variation (by replacing the Gaussian distributions in Eq.~\eqref{eq:spin-model} with Dirac delta functions), as in Refs.~\cite{Ng:2019jsx,Ng:2020ruv,Fernandez:2019qbj,Cheng:2022jsw,Ning:2026ebu}, we obtain a qualitatively different result: the inferred $m_b$ posterior exhibits a region of vanishing support that is broader than the exclusion obtained from individual high-spin events~\cite{Aswathi:2025nxa,Caputo:2025oap,LIGOScientific:2025brd,LIGOScientific:2026pwx}. As detailed in the End Matter, this excluded region
collapses when the single high-spin event GW241011 is removed, recovering a peak at $m_b\sim 10^{-12}\;\rm eV$.


Our approach assumes that the presence of a superradiance cloud does not bias the parameter estimation of each individual BBH event. 
Strikingly, however, the mass range identified in this study largely coincides with an analysis from an entirely independent method: Ref.~\cite{Roy:2025qaa} fits the waveform of GW190728 for the imprint of a scalar cloud on the binary dynamics, finding tentative evidence for a scalar of mass $\sim 10^{-12}\,\rm eV$ based on a superradiance-informed prior. We have confirmed that this consistency remains robust even after excluding GW190728 from our catalog and re-analyzing our model (see End Matter). 
Furthermore, the fact that our results favor a smaller scalar mass $m_b$ than reported in~\cite{Roy:2025qaa} could be attributed to potential parameter biases in individual BBH events in the presence of boson clouds. Such biases tend to overestimate the BH mass, as noted in~\cite{Roy:2025qaa}, and consequently lead to a systematic reduction in the inferred scalar boson mass associated with these BHs. Given that these two results arise from distinct mechanisms, their coincidence warrants further study, including the construction of a modified BBH catalog that incorporates the effects of boson clouds. 

In this work, we restrict the analysis to scalar bosons, but the method carries over to the vector case, which differs only in the functional form of $\chi_{\rm SR}$~\cite{Siemonsen:2022yyf,May:2024npn}, and we leave a dedicated vector analysis to future work. We also assume that the ultralight boson is only minimally coupled to gravity. However, non-gravitational interactions can become significant if the couplings are of sufficient strength. Recent studies have considered, for instance, scalar self-interactions~\cite{Arvanitaki:2010sy,Baryakhtar:2020gao,Witte:2024drg}, axion-photon couplings~\cite{Ikeda:2018nhb}, and kinetic mixing between dark photons and Standard Model photons~\cite{Caputo:2021efm,Siemonsen:2022ivj}.
These effects generally inhibit the growth of the superradiant cloud and suppress the efficiency of spindown through energy dissipation via bosonic and electromagnetic radiation.
As long as such couplings remain subdominant compared to the gravitational potential of the BH, the superradiant process remains largely unchanged. Exploring the parameter space of various non-gravitational interaction models lies beyond the scope of this study. 

For an ultralight boson with $m_b\simeq 10^{-12}\,\mathrm{eV}$, the associated quasi-monochromatic GW has a frequency $f_{\rm GW}\sim m_b/\pi\sim484\,\mathrm{Hz}$. Such signals could be observed by current and future ground-based GW detectors, either as resolvable narrow-band features or as part of a stochastic GW background~\cite{Brito:2017wnc}. 
Moreover, the coherent oscillation frequencies of such bosons lie within the audio band of ground-based interferometers. If these fields constitute the dark matter and couple to the Standard Model particles, they induce time-varying oscillations in both the sizes and optical properties of the instrument elements, producing an effective optical-path signal in laser interferometers and thereby giving rise to distinctive differential strains~\cite{Stadnik:2015xbn,Grote:2019uvn}. A recent search based on early O4 LVK data has extended these direct-detection limits to scalar, vector, and tensor fields over the $10^{-14}$–$10^{-11}\,\mathrm{eV}$ range~\cite{LIGOScientific:2025ttj}.
Future sensitivity can improve through longer integration, cross-correlation among detectors, and optimized material or geometric asymmetries~\cite{Grote:2019uvn}.  Resonant-mass detectors and atom-interferometric gravitational-wave detectors provide complementary future probes with different systematics~\cite{Arvanitaki:2015iga,Geraci:2016fva,Arvanitaki:2016fyj,MAGIS-100:2021etm}.

We emphasize that caution is needed when interpreting our results from the perspective of model selection. Since the Bayes factor inevitably depends on the choice of prior, comparisons between our results and other astrophysical models with empirical mass–spin correlations based solely on Bayes factors should be made with care. The future O5 observing run with improved sensitivity is expected to provide stronger evidence for model selection.

\begin{acknowledgments}

\noindent\textbf{\textit{Data Availability}} -- The  hierarchical Bayesian population-inference dataset that supports the findings of this article is publicly available~\cite{kou_2026_22242056}.

\noindent\textbf{\textit{Acknowledgements}} -- We would like to thank T\"ore Boybeyi, Michael Coughlin, Minyuan Jiang, Mario Spera and Salvatore Vitale for valuable discussions and comments. We are especially grateful to William East for conducting the internal collaboration review that helped improve the quality of the manuscript. R.D. would like to thank Minyuan Jiang, Yongcheng Wu and Lei Wu for their hospitality in hosting him at Nanjing Normal University. The authors acknowledge the computational resources provided by the Minnesota Supercomputing Institute (MSI). We are grateful for computational resources provided by the LIGO Laboratory and supported by National Science Foundation Grants PHY-0757058 and PHY-0823459. X.K. and V.M. are supported in part by the NSF grant PHY-2409173.  C.T. is supported by the National Natural Science Foundation of China (Grants No. 12405048). X.K. was partially supported by the Robert O. Pepin Fellowship at the University of Minnesota. This material is based upon work supported by NSF's LIGO Laboratory which is a major facility fully funded by the National Science Foundation.

\end{acknowledgments}

\bibliography{Ref.bib}

\clearpage

\begin{center}
{\large\bfseries End Matter}
\end{center}

\setcounter{section}{0}
\setcounter{subsection}{0}

\setcounter{secnumdepth}{2}

\section{Results without high or low spin events}
\begin{figure}[h]
    \centering
    \includegraphics[width=1.0\linewidth]{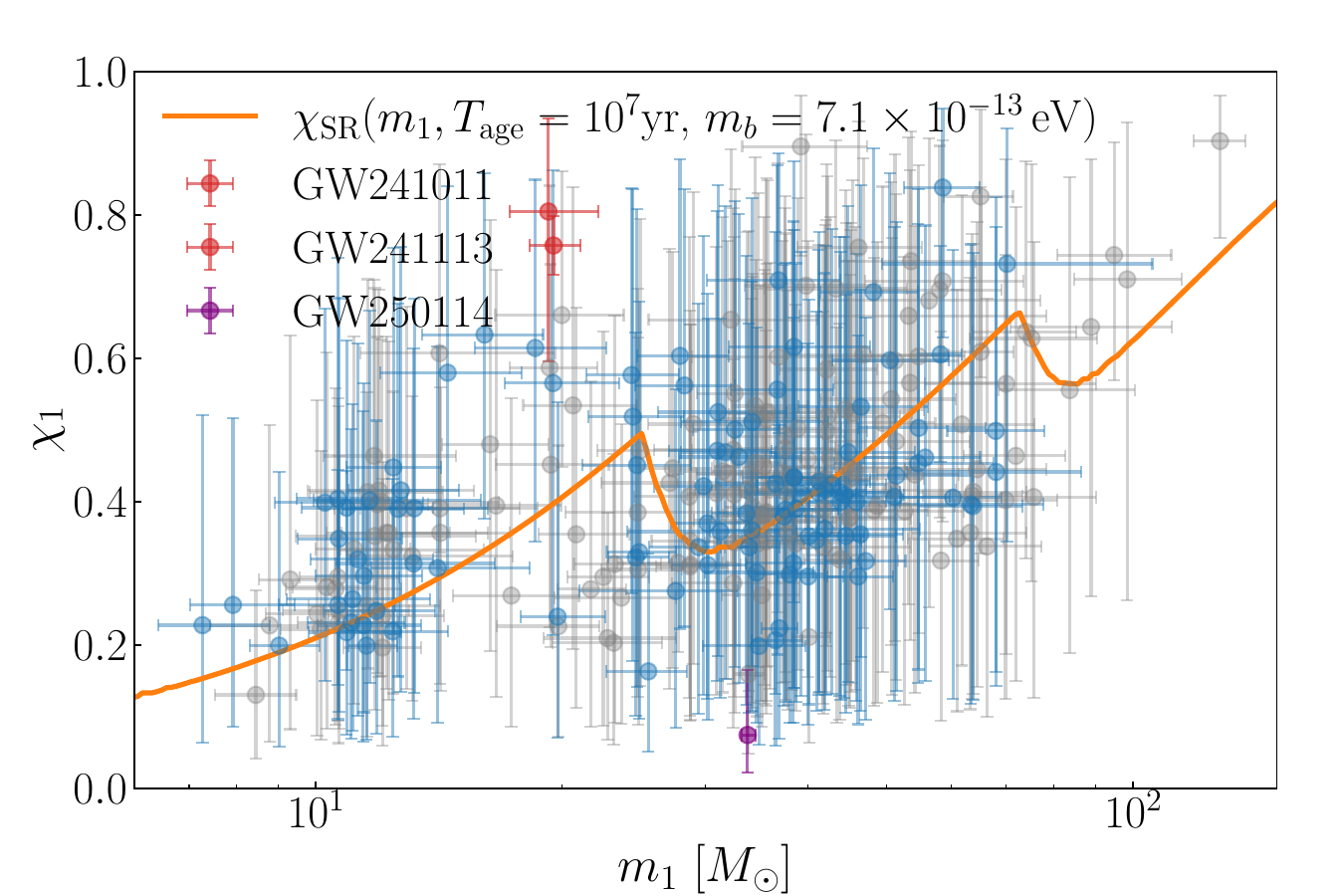}
    \caption{Mass-spin scatter plot of the BBH events from GWTC-5.0, with events already included in GWTC-4.0 shown in grey and those newly added in GWTC-5.0 in blue. Points represent the median values of the primary BH in binaries, with error bars denoting the $1\sigma$ uncertainties from the official LVK parameter estimation samples. The orange curve is a representative $\chi_{\rm SR}(m_1)$ trajectory at the full GWTC-5.0 posterior median $m_b=7.1\times10^{-13}\,\rm eV$  and fixed $T_{\rm age} = 10^{7}\,{\rm yr}$. It is included for illustration only, as the inference marginalizes over $T_{\rm age}$ per event. Three events singled out by dedicated LVK analyses are highlighted: the high-median-spin events GW241011 and GW241113 in red, and the low-spin event GW250114 in purple.}
    \label{fig:figM1}
\end{figure}

In Fig.~\ref{fig:figM1}, we show the mass-spin distribution of the BBH events included in our analysis, together with a representative $\chi_{\rm SR}$ trajectory at fixed $T_{\rm age}$. The events inherited from GWTC-4.0 (grey) are broadly consistent with this trajectory, as are most of those newly added in GWTC-5.0 (blue). Three exceptions are GW241011~\cite{LIGOScientific:2025brd}, GW241113~\cite{LIGOScientific:2026pwx}, and GW250114~\cite{LIGOScientific:2025rid,LIGOScientific:2025wao,Chandra:2025jfc}, each of which has been the subject of a dedicated LVK analysis.

\begin{figure}[h]
    \centering
    \includegraphics[width=0.9\linewidth]{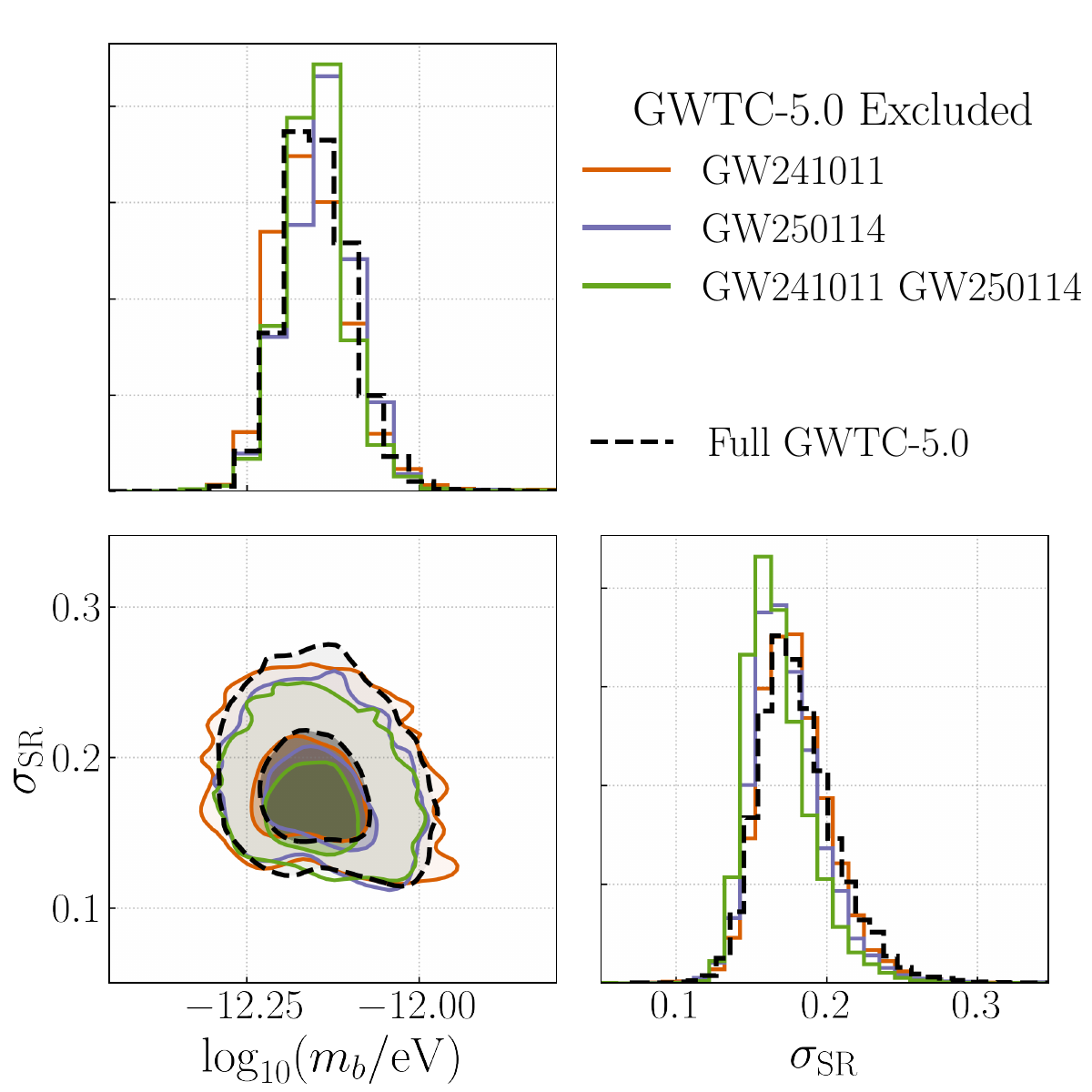}
    \caption{Posterior distributions for the boson mass $m_b$, the environment-induced spin dispersion $\sigma_{\rm SR}$ from GWTC-5.0 after some loud events are excluded. 
    Contours enclose $68\%$ and $95\%$ credible regions. }
    \label{fig:figM2}
\end{figure}

GW241011 and GW241113 have high median spins and lie above the trajectory, while GW250114 lies well below it. Their spins are not, however, equally well measured: GW250114 and GW241011 are among the loudest events ever detected, whereas GW241113 has a far lower SNR and correspondingly larger spin uncertainty. We therefore repeat the full inference with GW241011 and GW250114 removed (Table~\ref{tab:t1} and Fig.~\ref{fig:figM2}). These exclusions do not
create the preference for the superradiance spin model: $m_b$ is essentially unchanged, while the increase in $\ln\mathcal{B}$ quantifies the modest evidence penalty associated with accommodating these two loud events that depart from the trajectory, which are absorbed into the full-catalog fit through a slightly larger $\sigma_{\rm SR}$. Excluding GW241113 yields no improvement in $\ln\mathcal{B}$, consistent with the weaker constraint it places on the boson mass in Ref.~\cite{LIGOScientific:2026pwx}.

\section{Results without GW190728}

\begin{figure}[h]
    \centering
    \includegraphics[width=0.9\linewidth]
    {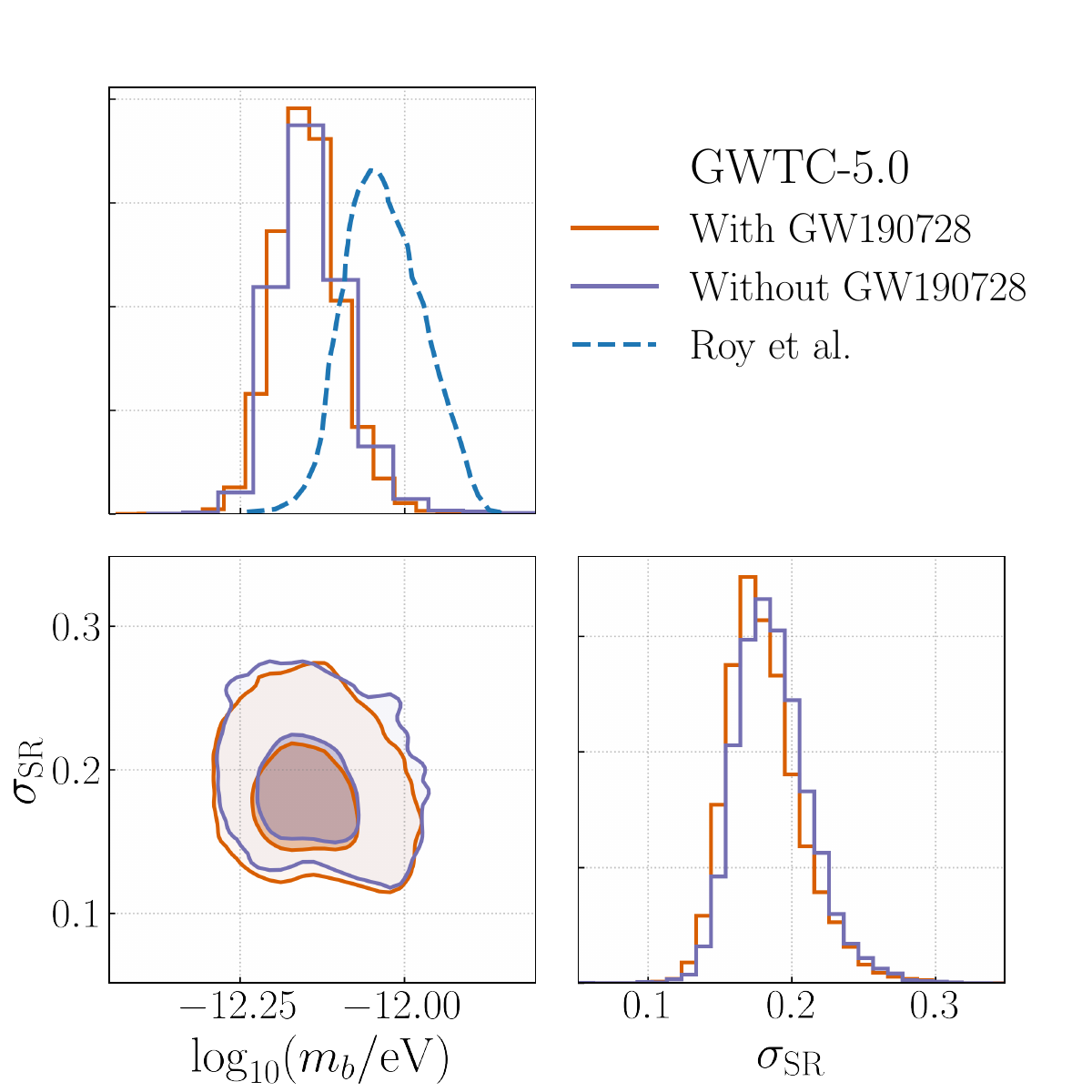}
    \caption{Posteriors on the scalar boson mass $m_b$ and the environment-induced spin dispersion $\sigma_{\rm SR}$ from the GWTC-5.0 catalogs with (orange) and without (purple) the GW190728 event. Contours show $68\%$ and $95\%$ credible regions. The boson mass $m_b$ from the waveform analysis of GW190728 by Roy et al.~\cite{Roy:2025qaa} is indicated by the blue dashed curve.}
    \label{fig:figM4}
\end{figure}

In Fig.~\ref{fig:figM4}, we present the posteriors on the boson mass $m_b$ and the spin dispersion $\sigma_{\rm SR}$ inferred from the GWTC-5.0 with and without the inclusion of the GW190728 event. The posterior for the scalar boson mass remains consistent with that reported by Roy et al.~\cite{Roy:2025qaa}, even after removing GW190728, which is the basis of their analysis, from the catalog.

\section{Results of delta function spin model}

In this section, we present the results of a superradiance model without considering any environment-induced spin  variations, and show that this can lead to qualitatively different results. In this context, the spin magnitude model can be written as 
\begin{align}
\label{eq:spin-model-delta}
&p(\chi\,|\,M_f, T_{\rm age}, m_b, \Lambda_s) \\
&\qquad= \pi_{\rm base}(\chi\,|\,\Lambda_s)\,\Theta(\chi_{\rm SR}-\chi) + C\,\delta(\chi - \chi_{\rm SR})\,,\nonumber
\end{align}
where $\delta(\chi)$ denotes the Dirac delta function. This spin model, studied in Ref.~\cite{Ng:2019jsx,Ng:2020ruv,Fernandez:2019qbj,Cheng:2022jsw,Ning:2026ebu}, is consistent with the model used in boson mass constraints from individual events~\cite{Aswathi:2025nxa,Caputo:2025oap,LIGOScientific:2025brd,LIGOScientific:2026pwx}, and is hereafter referred to as the \textit{delta function spin model}. For all runs in this section we choose the baseline model to be a beta distribution and fix $T_{\rm age} = 10\,\rm{Myr}$, matching the choice made in Refs.~\cite{Ng:2019jsx,Ng:2020ruv,Fernandez:2019qbj,Cheng:2022jsw,Ning:2026ebu}. Although Ref.~\cite{Ning:2026ebu} instead marginalizes over $T_{\rm age}$ at the population level, they report that fixing it to this value does not appreciably alter their results.  However, the infinitesimally sharp peak in the distribution function will make any direct inference impossible, since the standard Monte Carlo estimator evaluates the population model at discrete posterior samples (see Supplemental Material): a term with support on a set of zero measure contributes nothing, while narrowly smoothing it inflates the variance of the likelihood estimator so severely that the variance
cut removes a large region of parameter space unphysically. Consequently, one has to work with the spin distribution after analytically integrating Eq.~\eqref{eq:spin-model-delta}, though only for the delta-function term.

Following Refs.~\cite{Ng:2019jsx,Ng:2020ruv,Fernandez:2019qbj,Cheng:2022jsw,Ning:2026ebu}, we restrict this section to the mass and spin-magnitude sector and adopt the same single power-law mass function and uniform mass ratio distribution used in those
analyses. Assuming that the single-event posterior
factorizes between the mass and spin sectors, $p(m_i,\chi_i\,|\,d) \approx p(m_i\,|\,d)\,p(\chi_i\,|\,d)$, and using the flat PE prior on spin magnitude, the delta-function contribution to the per-event marginal likelihood reduces to
\begin{equation}
\label{eq:delta_spin_sample}
\frac{1}{N_{\rm PE}}\sum_{j=1}^{N_{\rm PE}}
\frac{\pi(m_{i,j}\,|\,\Lambda_m)}{\pi_{\rm PE}(m_{i,j})}\,
C_j\, p(\chi^{j}_{{\rm SR},i}\,|\,d),
\end{equation}
in which $N_{\rm PE}$ is the number of posterior samples, $i=1,2$ labels the binary component and $j$ indexes the posterior sample. The normalization $C_j$ of Eq.~\eqref{eq:delta_spin_sample} is evaluated per sample, since it depends on $\chi^j_{{\rm SR},i}$. The Monte Carlo sum runs only over the mass sector and the spin integral has been evaluated exactly, leaving the marginal spin posterior density evaluated on the $\chi_{\rm SR}$ trajectory. Evaluating the delta-function term therefore requires the posterior density itself rather than posterior samples, and we obtain $p(\chi^{j}_{{\rm SR},i}\,|\,d)$ by kernel density estimation of the spin samples for each event. This treatment necessarily neglects all correlations between spin magnitude and the other source parameters, particularly the spin-orbit tilt angles, and treats $\chi_1$ and $\chi_2$ as mutually independent. Refs.~\cite{Hussain:2025llf, Hussain:2026pfm} instead construct a truncated Gaussian mixture representation of the full single-event posterior, retaining these correlations at the cost of a more expensive fitting step. 
Extending that framework to the superradiance spin models is a natural next step.

\begin{figure}[t]
    \centering
     \includegraphics[width=1\linewidth]{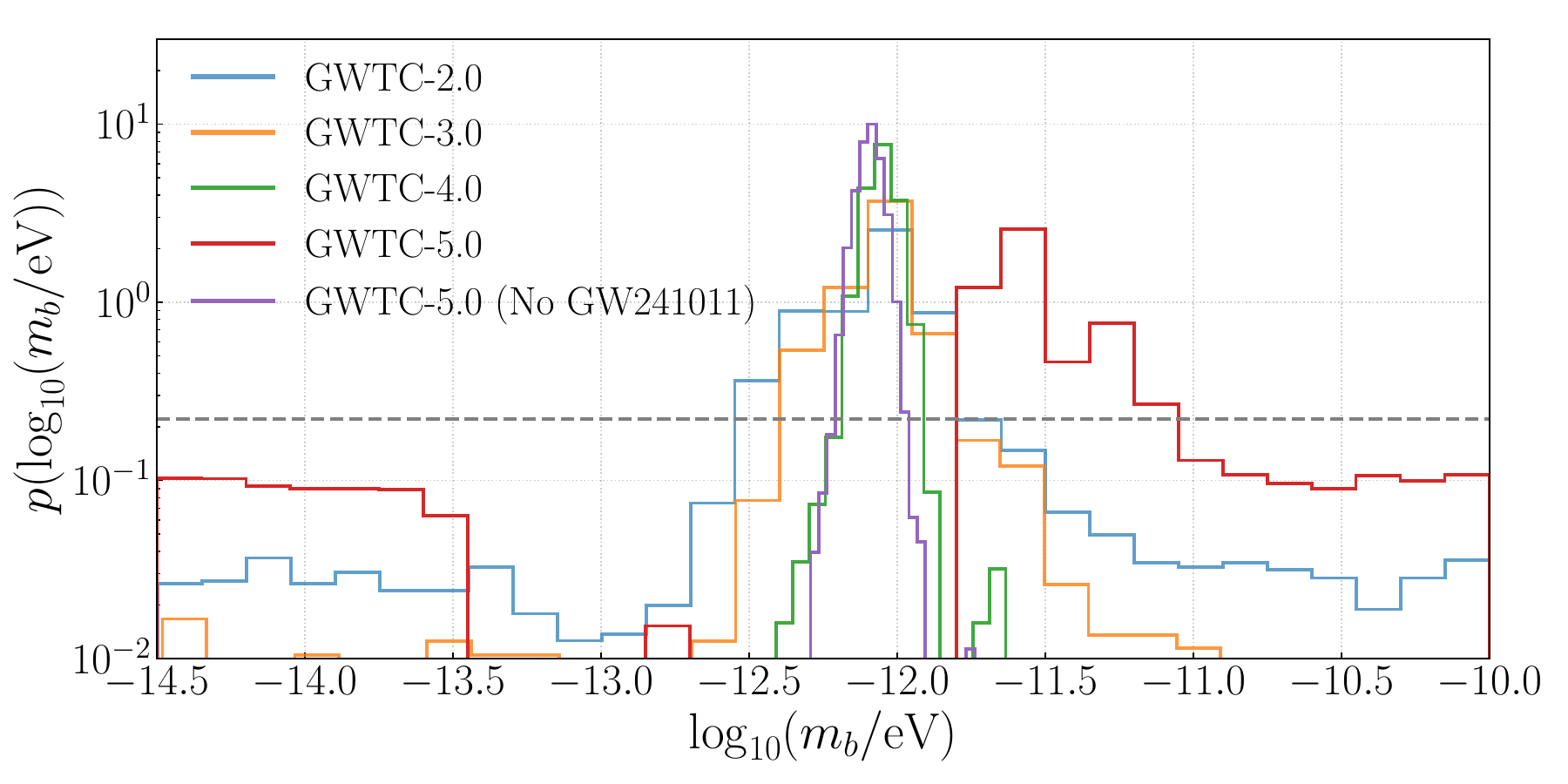}
    \caption{Posteriors on the scalar boson mass $m_b$ from the GWTC-2.0 to GWTC-5.0 catalogs, based on the \textit{delta function spin model} with $T_{\rm age} = 10\,\rm{Myr}$ fixed for all events. The grey dashed line represents prior of $m_b$ in the log scale.}
    \label{fig:figM5}
\end{figure}

We present the posteriors for $m_b$ based on the \textit{delta function spin model} from GWTC-2.0 through GWTC-5.0 in Fig.~\ref{fig:figM5}. The result from GWTC-2.0 is consistent with that of~\cite{Ng:2020ruv}. A narrower peak in $m_b$ becomes clearly visible when considering GWTC-3.0 through GWTC-4.0, consistent with the result given by the extended \textit{superradiance spin model}. When the full GWTC-5.0 dataset is considered, this peak broadens substantially and the posterior develops an appreciable region of vanishing support, approximately between $[3.5\times10^{-14},\,1.6\times10^{-12}]\,\rm eV$,
consistent with Ref.~\cite{Ning:2026ebu} and broader than the exclusion obtained from individual high-spin events~\cite{Aswathi:2025nxa, Caputo:2025oap,LIGOScientific:2025brd,LIGOScientific:2026pwx}.
However, when the single high-spin event GW241011 is excluded, the peak is recovered, suggesting that the excluded region arises solely from this event.


\clearpage
\onecolumngrid

\begin{center}
{\large\bfseries Supplemental Material}
\end{center}

\setcounter{section}{0}
\setcounter{subsection}{0}
\setcounter{equation}{0}
\setcounter{figure}{0}
\setcounter{table}{0}

\setcounter{secnumdepth}{2}

\renewcommand{\thesection}{S\arabic{section}}
\renewcommand{\thesubsection}{\thesection.\arabic{subsection}}
\renewcommand{\theequation}{S\arabic{equation}}
\renewcommand{\thefigure}{S\arabic{figure}}
\renewcommand{\thetable}{S\arabic{table}}

\setcounter{page}{1} 

\section{Hierarchical Bayesian Inference and marginalization of \texorpdfstring{$T_{\rm age}$}{Tage}} \label{sup:hierarchical}

We aim to infer the population hyperparameters $\Lambda$, which include the boson mass $m_b$, the environment-induced spin dispersion $\sigma_{\rm SR}$, and the baseline model parameters, from a catalog of $N_{\rm det}$ BBH events $\{d_i\}$ using the rate-marginalized hierarchical likelihood~\cite{Mandel:2018mve, Thrane:2018qnx, Vitale:2020aaz}
\begin{equation} \label{eq:hierarchical_likelihood}
    \mathcal{L}(\{d_i\}\,|\,\Lambda)\propto\prod^{N_{\rm det}}_{i=1}\frac{\int\mathcal{L}(d_i\,|\,\theta)\pi(\theta\,|\,\Lambda)d\theta}{\xi(\Lambda)},
\end{equation}
where $\theta$ denotes the single-event source parameters, $\mathcal{L}(d_i\,|\,\theta)$ is the single-event likelihood, and $\pi(\theta\,|\,\Lambda)$ is the population model specifying the astrophysical distribution of source parameters conditional on the hyperparameters $\Lambda$. This form assumes statistically independent events drawn from an inhomogeneous Poisson process, whose overall rate has been analytically marginalized under a $1/N$ prior~\cite{Fishbach:2018edt,Essick:2023upv}. The selection function $\xi(\Lambda)$ is the detectable fraction of the population under the adopted detection criterion and accounts for survey selection bias against weakly emitting sources.

In our analysis, the superradiance signature enters the population model by depleting the high-spin population above the mass-dependent threshold $\chi_{\rm SR}(M_f,T_{\rm age},m_b)$. Because $\chi_{\rm SR}$ also depends on the formation-to-merger interval $T_{\rm age}$, which is not directly measurable for any individual BBH event and may vary across the catalog depending on each binary’s formation history, we treat it as a per-event intrinsic source parameter and marginalize over it independently for each event using the log-uniform prior $\pi(T_{\rm age})$ between $1\,\rm Myr$ and $2.2\,\rm Gyr$ motivated in the main text. We derive the explicit form of this marginalization below.

Assuming that $T_{\rm age}$ is uncorrelated with the other source parameters for population-level parameter inference, the augmented population model factorizes as
\begin{equation} \label{eq: joint_prior}
\pi(\theta, T_{\rm age}\,|\, \Lambda) = \pi(\theta\,|\,\Lambda, T_{\rm age})\pi(T_{\rm age}).
\end{equation}
The per-event integral in Eq.~\eqref{eq:hierarchical_likelihood} is generalized to
\begin{equation} \label{eq:event_integral}
\int \mathcal{L}(d_i\,|\,\theta, T_{\rm age})\pi(\theta, T_{\rm age}\,|\, \Lambda)\,d\theta\,dT_{\rm age}.
\end{equation}
Because the size of $T_{\rm age}$ does not change the GW waveform from detected BBHs, we have $\mathcal{L}(d_i\,|\, \theta, T_{\rm age}) = \mathcal{L}(d_i\,|\, \theta)$, and the joint event-level posterior factorizes as
\begin{equation} \label{eq:joint_event_posterior}
p(\theta, T_{\rm age}\,|\, d_i) = \mathcal{L}(d_i\,|\,\theta, T_{\rm age})\pi_{\rm PE}(\theta)\pi(T_{\rm age}) = p(\theta\,|\, d_i)\pi(T_{\rm age})\,,
\end{equation}
where $\pi_{\rm PE}(\theta)$ is the default parameter-estimation prior adopted in the GWTCs analysis~\cite{LIGOScientific:2026exk,LIGOScientific:2026wfs}. Eq.~\eqref{eq:joint_event_posterior} shows that the joint posterior decomposes into the standard event-level posterior $p(\theta\,|\, d_i)$ multiplied by the independent prior on $T_{\rm age}$. The existing posterior samples in GWTCs therefore remain valid as samples from $p(\theta\,|\, d_i)$, and the joint posterior can be sampled simply by appending an independent draw $T_{\mathrm{age}, j}$ from $\pi(T_{\rm age})$ to each sample $j$. These samples are taken from the public LVK data products for each catalog. For events in O4b~\cite{LIGOScientific:2026wfs} we use the \texttt{IMRPhenomXPHM\_SpinTaylor} samples~\cite{Pratten:2020ceb, Colleoni:2024knd}.
For O4a events~\cite{LIGOScientific:2026tep} we use \texttt{NRSur7dq4}
samples~\cite{Varma:2019csw} where available, and otherwise the \texttt{Mixed} samples combining \texttt{IMRPhenomXPHM\_SpinTaylor} and
\texttt{SEOBNRv5PHM}~\cite{Ramos-Buades:2023ehm, Pompili:2023tna}. For events prior to O4a we use the \texttt{Mixed} samples released with GWTC-3.0~\cite{KAGRA:2021vkt} and GWTC-2.1~\cite{LIGOScientific:2021usb}.

Substituting Eq.~\eqref{eq: joint_prior} into Eq.~\eqref{eq:event_integral}, the integral can be evaluated by Monte Carlo importance sampling~\cite{Tiwari:2017ndi, Farr:2019rap} as
\begin{equation} \label{eq:mc_event}
\begin{aligned}
\int \mathcal{L}(d_i\,|\, \theta, T_{\rm age})\pi(\theta\,|\,\Lambda, T_{\rm age})\pi(T_{\rm age})d\theta dT_{\rm age} &\approx \frac{1}{N_{\rm PE}}\sum^{N_{\rm PE}}_{j=1}\frac{\pi(\theta_{i,j}\,|\,\Lambda, T_{\mathrm{age}, j})\pi(T_{\mathrm{age}, j})}{\pi_{\rm PE}(\theta_{i,j})\pi(T_{\mathrm{age}, j})}\\
&= \frac{1}{N_{\rm PE}}\sum^{N_{\rm PE}}_{j=1}\frac{\pi(\theta_{i,j}\,|\,\Lambda, T_{\mathrm{age}, j})}{\pi_{\rm PE}(\theta_{i,j})},
\end{aligned}
\end{equation}
where $\{\theta_{i,j}\}^{N_{\rm PE}}_{j=1}$ are $N_{\rm PE}$ samples from the posterior of the $i$th event $d_i$, and each is paired with an independent draw $T_{\mathrm{age}, j} $ from $ \pi(T_{\rm age})$.

The selection function is evaluated by a similar Monte Carlo procedure on the LVK injection set~\cite{Essick:2025zed}. Again, each recovered injection $\theta_j$ is assigned an independent draw $T_{\mathrm{age},j}$ from $\pi(T_{\rm age})$, and
\begin{equation} \label{eq:mc_selection}
\xi(\Lambda) \approx \frac{1}{N_{\rm draw}}\sum^{N_{\rm found}}_{j=1}\frac{\pi(\theta_j\,|\, \Lambda, T_{\mathrm{age},j})}{\pi_{\rm draw}(\theta_j)},
\end{equation}
where $\pi_{\rm draw}$ is the injection sampling distribution and the sum runs over the $N_{\rm found}$ injections recovered above the detection threshold. This ensures that the population model and the selection function are evaluated consistently under the same per-event treatment of $T_{\rm age}$.

\begin{figure}[t]
    \centering
    \includegraphics[width=0.49\linewidth]{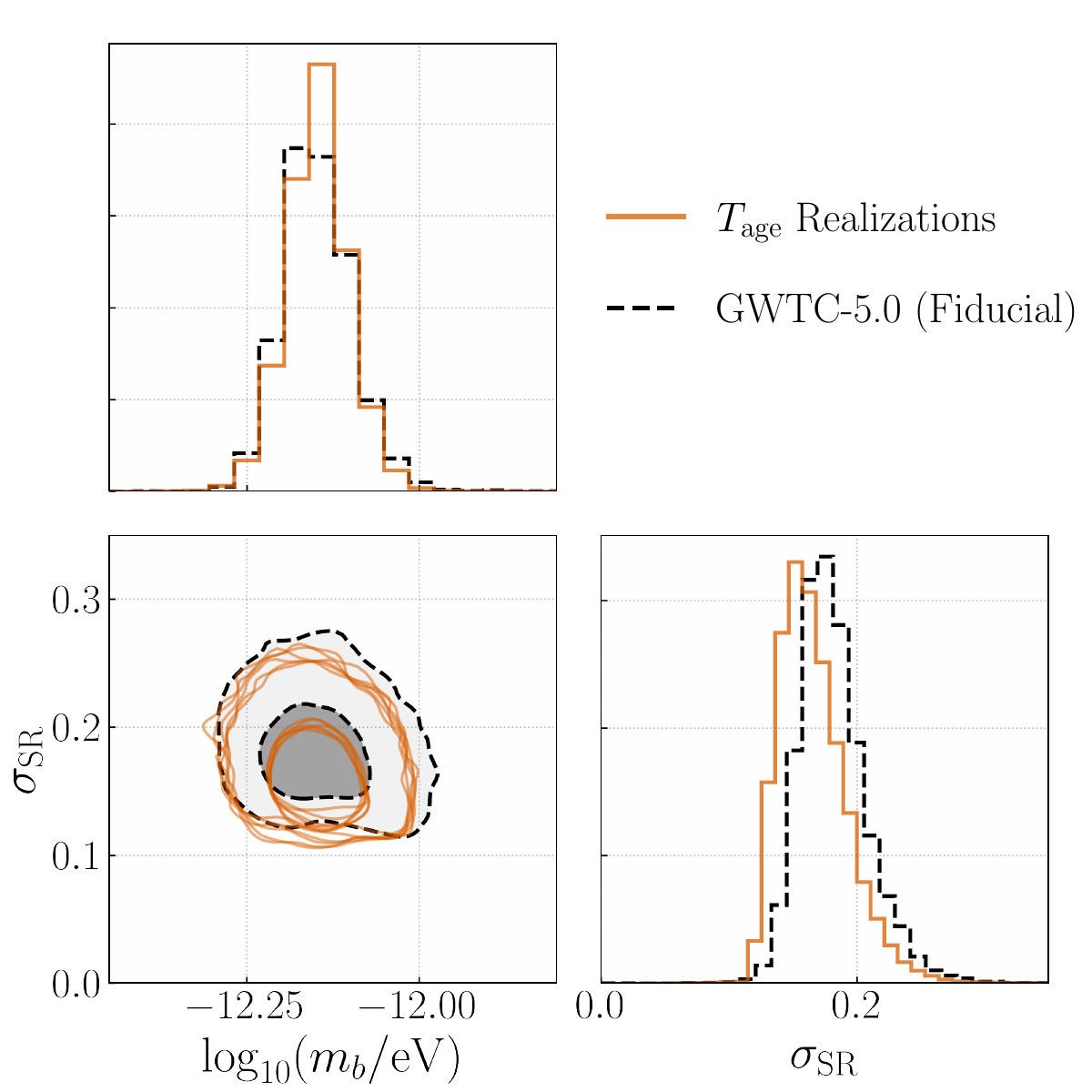}
    \includegraphics[width=0.49\linewidth]{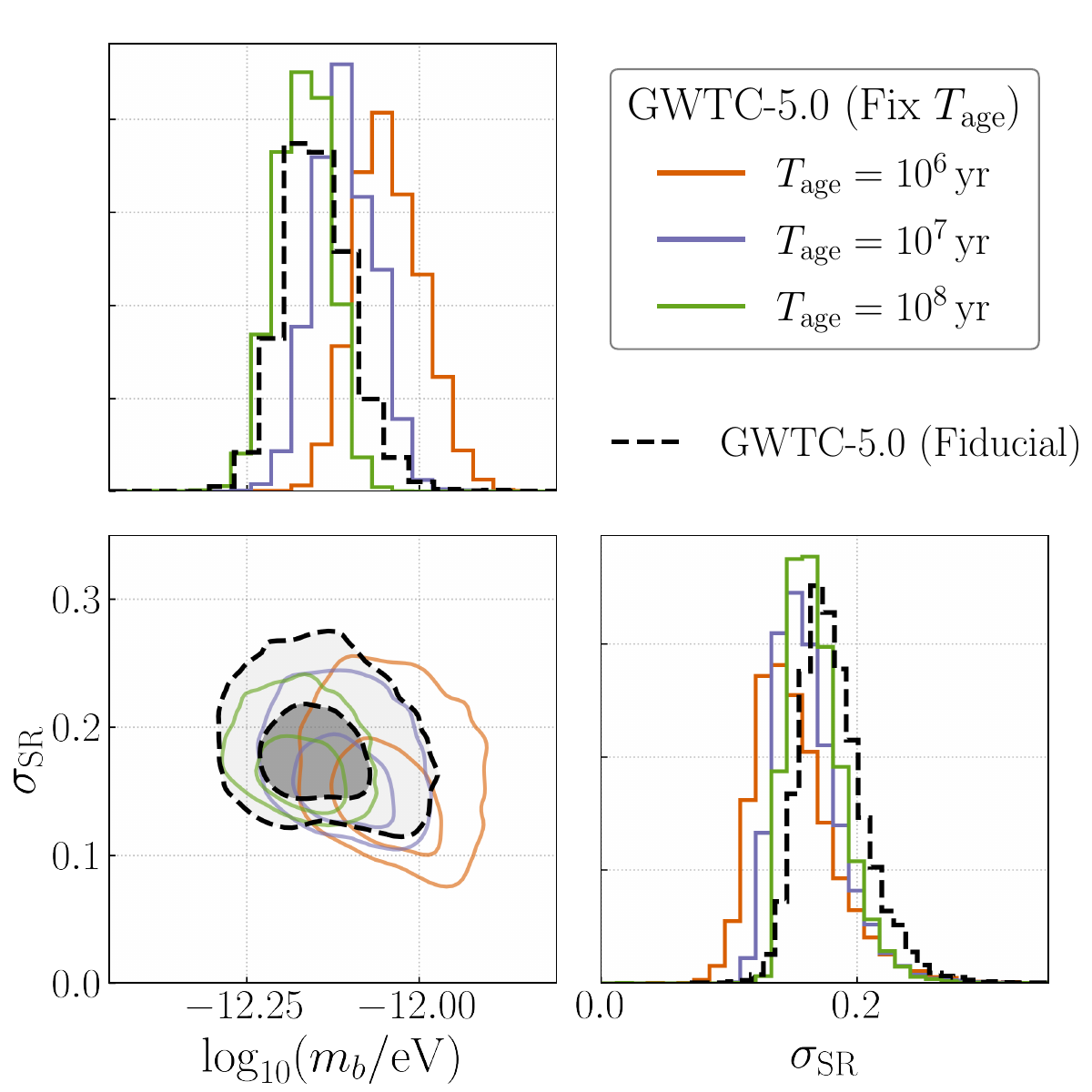}
    \caption{Posteriors on the superradiance parameters $\{m_b, \sigma_{\rm SR}\}$ inferred from GWTC-5.0, with the mass-function hyperparameters held fixed at their posterior medians from the baseline analysis of Ref.~\cite{LIGOScientific:2026ctl}. Left: six independent realizations of drawing from $\pi(T_{\rm age})$ (orange), where diagonal panels show the marginalized one-dimensional distributions obtained by combining all realizations. Right: three runs, each with $T_{\rm age}$ fixed to a common value of $10^{6}$, $10^{7}$, or $10^{8}\,{\rm yr}$ for all events. The fiducial result of Fig.~\ref{fig:fig1}, in which all hyperparameters are sampled, is overlaid in dashed black for reference. Contours show $68\%$ and $95\%$ credible regions.}
    \label{fig:figS1}
\end{figure}

The accuracy of these Monte Carlo estimators has a direct impact on hierarchical inference and must therefore be monitored~\cite{Farr:2019rap, Essick:2022ojx, Talbot:2023pex, Heinzel:2025ogf}. Denoting the summands of Eqs.~\eqref{eq:mc_event} and \eqref{eq:mc_selection} by $w_j$, the effective number of contributing samples is
\begin{equation} \label{eq:neff}
N_{\rm eff} = \frac{\left(\sum_j w_j\right)^2}{\sum_j w_j^2},
\end{equation}
evaluated separately for each per-event sum and for the injection sum. Following Ref.~\cite{Essick:2022ojx}, we require $N^{\rm inj}_{\rm eff} \geq 4 N_{\rm det}$ for the selection function, and we additionally require $\min_i N^{\rm event}_{{\rm eff},i} \geq 20$ across the per-event sums. Rather than imposing these thresholds as hard cuts, which would introduce a discontinuity in the likelihood surface, we follow Ref.~\cite{Callister:2023tgi} and add penalties to the log-likelihood to exclude poorly sampled regions of the parameter space,
\begin{equation} \label{eq:penalty}
\ln S\!\left(\frac{N^{\rm inj}_{\rm eff}(\Lambda)}{4N_{\rm det}}\right)
+ \ln S\!\left(\frac{\min _i N^{\rm event}_{{\rm eff},i}}{20}\right), \qquad
S(x) = \frac{1}{1+x^{-30}},
\end{equation}
where $S(x)$ approaches  unity when the corresponding criterion is
satisfied, while smoothly suppressing the likelihood as the boundary is approached from above. 

The second term guards against individual events for which the per-event sums are poorly converged, an issue that would not be detected by the aggregate condition on $N^{\rm inj}_{\rm eff}$ alone. We adopt a stricter threshold here than that used in Ref.~\cite{Callister:2023tgi,Antonini:2024het,Antonini:2025ilj,Antonini:2025zzw,LIGOScientific:2026ctl,Flanagan:2026ayy}, because the pile-up feature of the \textit{superradiance spin model} concentrates population support in a narrow region of the spin distribution and can leave individual events supported by very few effective samples.

The tests in this section examine two separate consequences of our treatment of $T_{\rm age}$: the variance introduced by the random draws themselves, and the sensitivity of the inference to the assumed values of $T_{\rm age}$. To control computational cost, all runs shown in Fig.~\ref{fig:figS1} hold the mass-function hyperparameters fixed at their posterior medians from the baseline analysis of Ref.~\cite{LIGOScientific:2026ctl}; we have verified that the \textit{superradiance spin
model} leaves the inferred mass function negligibly changed relative to the baseline model. The fiducial result, in which all hyperparameters are sampled, is overlaid for
reference. The offset between it and the fixed mass function runs is small but visible, reflecting the additional uncertainty propagated by marginalizing over
the mass function rather than fixing it; this is one reason we infer the mass, redshift, spin, and superradiance sectors jointly, whereas previous population
searches for superradiance have held the mass function
fixed~\cite{Ng:2019jsx,Ng:2020ruv,Fernandez:2019qbj,Cheng:2022jsw, Ning:2026ebu}.

Because $T_{\rm age}$ is appended to each posterior sample as a single draw from $\pi(T_{\rm age})$ rather than integrated analytically, different draws yield
slightly different Monte Carlo estimators of the hierarchical likelihood. The left panel of Fig.~\ref{fig:figS1} shows six independent realizations of drawing $T_{\rm age}$ values, applied to both the event posterior samples and the injection set. These runs are identical apart from the draws themselves, so the
spread among them isolates the realization variance, which is consistent with stochastic sampling noise and well below the statistical uncertainty on $m_b$ and
$\sigma_{\rm SR}$. The right panel instead fixes $T_{\rm age}$ to a common value for all events, at $10^{6}$, $10^{7}$, and $10^{8}\,{\rm yr}$. The inferred boson mass shifts monotonically toward smaller values as $T_{\rm age}$ increases, since a longer superradiance-active interval allows the instability to deplete spins efficiently at smaller $\alpha_g$. The shift is modest, spanning about $0.15$ dex over two decades in $T_{\rm age}$, comparable to the width of the individual posteriors, and $\sigma_{\rm SR}$ is largely insensitive to this choice. The fiducial result
lies within the range spanned by the three fixed values and is broader than any of them, as expected once the spread in formation-to-merger times is propagated into the posterior rather than fixed by assumption. This supports treating $T_{\rm age}$ as an event-specific parameter rather than imposing a single value
across the population, as done in Refs.~\cite{Cheng:2022jsw, Ning:2026ebu}.

\section{Validations on mock catalogs}

Our goal in this section is to generate mock catalogs and analyze them using our methodology to ensure that the inferred boson mass is not biased by a particular realization of the observed events or by the overdense region in the BH mass function. Starting from a known BBH population whose spin distribution follows our superradiance spin model, we generate independent simulated GW catalogs, construct realistic mock parameter-estimation posteriors for each detected event, and apply the full hierarchical inference pipeline, including per-event $T_{\rm age}$ marginalization, to the resulting samples.

Because our \textit{superradiance spin model} treats the primary and secondary spins separately, with each BH spin correlated with its own mass through $\chi_{\rm SR}$, the model’s central prediction can be tested, without loss of generality, using an equal-mass configuration with a single spinning BH. We therefore restrict the mock catalog to binaries with mass ratio $q=1$, $\chi_2=0$, and $\chi_1$ aligned with the orbital angular momentum. This reduces the validation to a one-dimensional test of the recovery of $\chi_{\rm SR}(m_1)$ from the $(m_1,\chi_1)$ distribution.

\begin{figure}[t]
    \centering
    \includegraphics[width=0.6\linewidth]{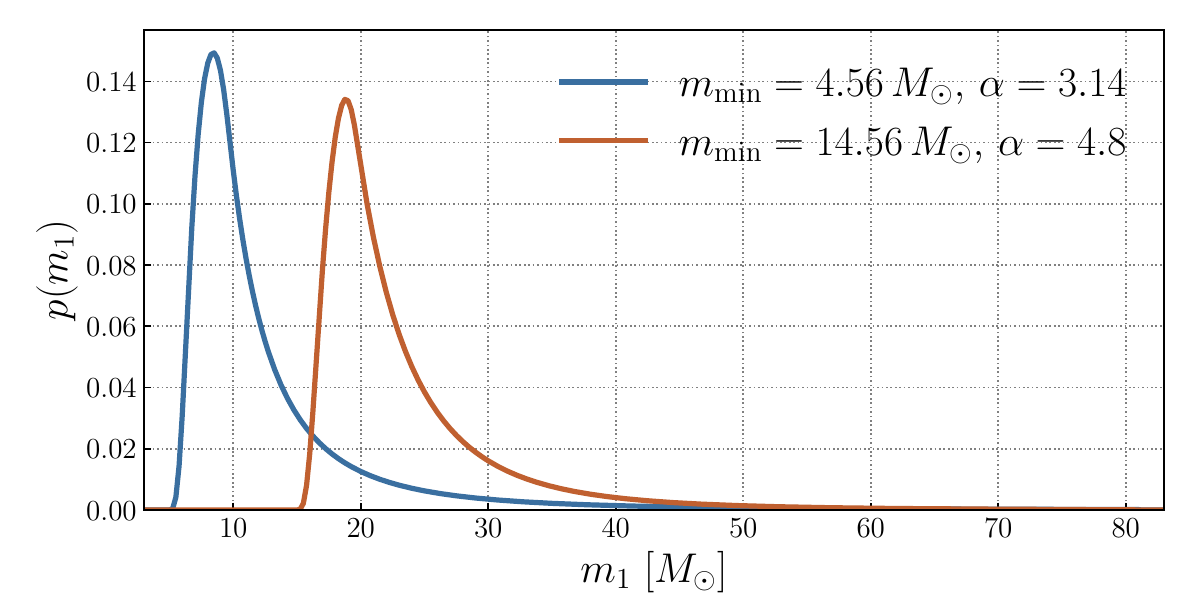}
    \caption{Injected mass functions $\pi(m)$ for the fiducial mock catalog (blue; $m_{\mathrm{min}}=4.56\, M_{\odot}, \alpha=3.14$) and the shifted mock catalog (brown; $m_{\mathrm{min}}=14.56\, M_{\odot}, \alpha=4.8$), with $\delta_m = 5.96\, M_{\odot}$ and $m_{\mathrm{max}}=81.08\, M_{\odot}$ held fixed. The two populations place the peak of $\pi(m)$ near $10\, M_{\odot}$ and $20\, M_{\odot}$, respectively.}
    \label{fig:figS2}
\end{figure}

The injected BBH population is modeled using the \textit{Power Law + Peak} mass model~\cite{KAGRA:2021duu}, with the Gaussian peak component disabled (mixing fraction $\lambda_{\rm peak}=0$), following the simplified mass-function prescription of~\cite{Farah:2023vsc}. The resulting distribution is a smoothed power law, in which the peak of $\pi(m)$ corresponds to the maximum of the probability density at the low-mass smoothing turn-on, controlled by the parameters $m_{\rm min}$ and $\delta_m$. We adopt fiducial mass-function parameters $m_{\mathrm{min}}=4.56\,M_{\odot}, m_{\mathrm{max}}=81.08\, M_{\odot}, \delta_m = 5.96\, M_{\odot}$, and power-law slope $\alpha=3.14$, which place the peak of $\pi(m)$ around $10\, M_{\odot}$, consistent with the low-mass structure observed in GWTC-5.0~\cite{LIGOScientific:2026ctl}; this distribution is shown in blue in Fig.~\ref{fig:figS2}, together with the
shifted mass function described below. This simpler parameterization, which contains fewer hyperparameters than the baseline \textit{Broken Power Law + 2 Peaks} model, is sufficient because the mock-data analysis focuses on the spin sector. The redshift evolution follows the standard \textit{Power Law Redshift model}, with $\kappa=3.2$. The spin sector adopts the  \textit{superradiance spin model} for scalars described in the main text, with injected hyperparameters $\log_{10}(m_b/\rm{eV})=-12.1$ and $\sigma_{\rm SR}=0.16$, and with the baseline spin hyperparameters set to $\mu_\chi = 0.6$ and $\sigma_\chi = 0.4$, the medians of our fiducial superradiance-extended inference on the full GWTC-5.0 catalog.


Because GW detectors are insensitive to $T_{\rm age}$, the catalog generation is divided into two steps. We first construct the population predictive distribution $\pi(m_1, \chi_1\,|\, \Lambda) = \int \pi(m_1,\chi_1\,|\, \Lambda, T_{\rm age})\pi(T_{\rm age})dT_{\rm age}$ by Monte Carlo marginalization over the same log-uniform $T_{\rm age}$ prior used in the main text, and draw the true source parameters: $(m_1, \chi_1)$ from this $T_{\rm age}$-marginalized distribution and the redshift $z$ from the redshift evolution model. Each draw is then assigned an independent $T^{i}_{\mathrm{age}}$ value from the same $\pi(T_{\rm age})$ prior. This assignment carries no observational information and does not affect detectability; rather, it constitutes event-level metadata used only at inference time, hence mirroring the per-event $T_{\rm age}$ treatment applied to the GWTC-3.0 to GWTC-5.0 catalogs.

\begin{figure}[h]
    \centering
    \includegraphics[width=0.5\linewidth]{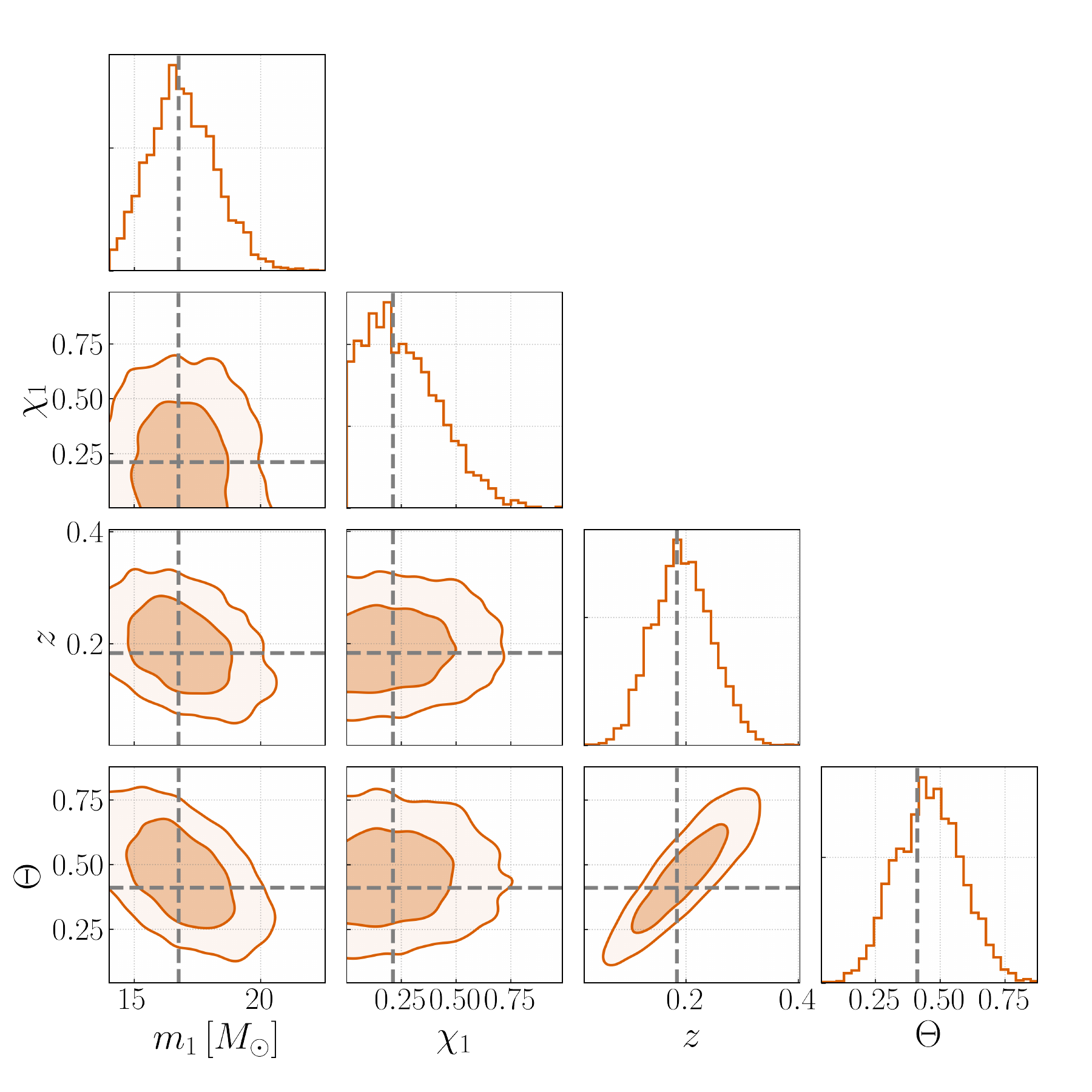}
    \caption{Posteriors for a representative detected event in the mock catalog, with observed signal-to-noise ratio $\rho_{\rm obs}=12.7$. Shown are the primary mass $m_1$ in the source frame, primary spin magnitude $\chi_1$, source redshift $z$, and sky-angle factor $\Theta$. Grey dashed lines mark the injected true values; black contours show the $68\%$ and $95\%$ credible regions.}
    \label{fig:figS3}
\end{figure}

\begin{figure}[t]
    \centering
    \includegraphics[width=0.49\linewidth]{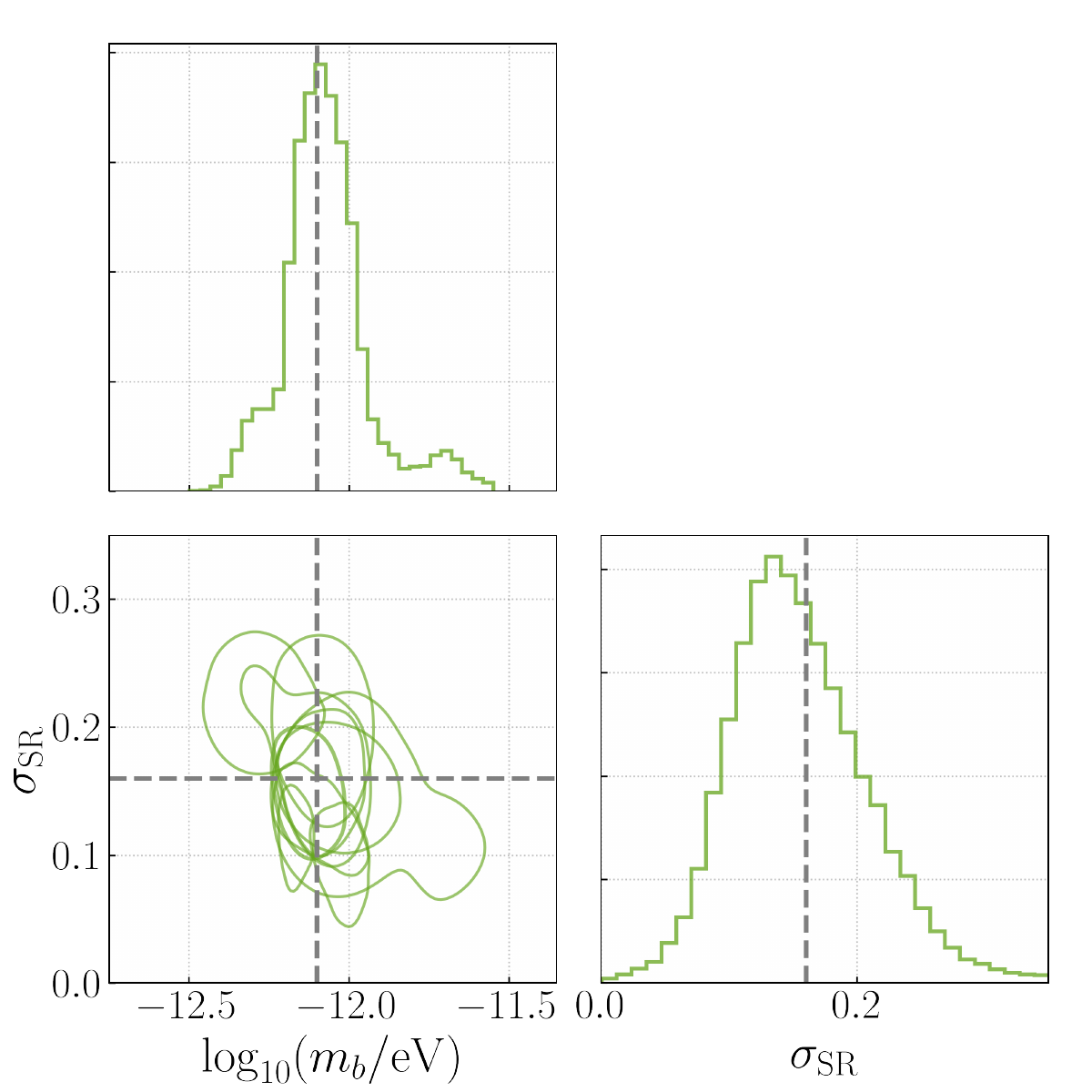}
    \includegraphics[width=0.49\linewidth]{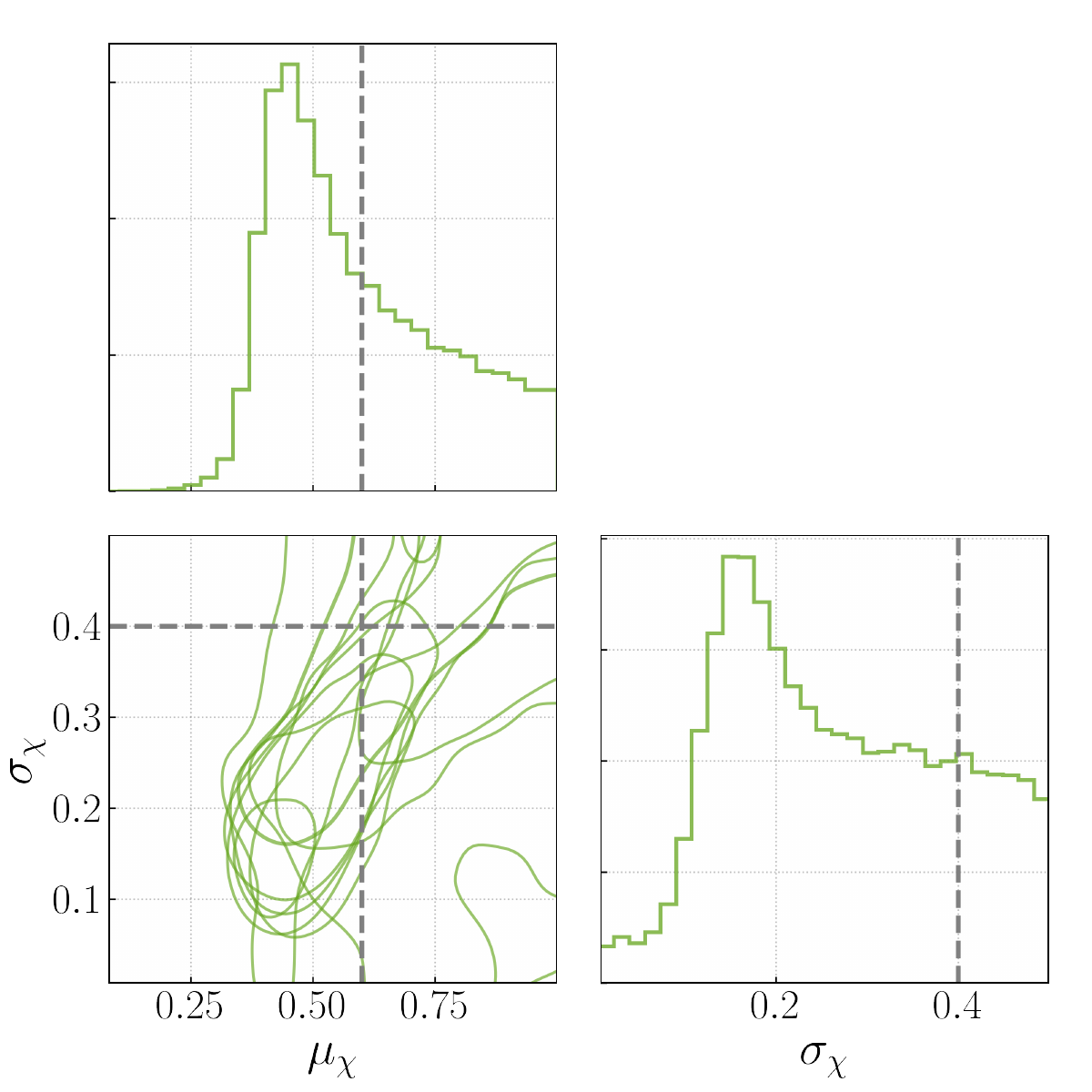}
    \caption{Reconstructed $68\%$ credible contours on the superradiance parameters $\{m_b, \sigma_{\mathrm{SR}}\}$ and the baseline natal-spin parameters $\{\mu_{\chi}, \sigma_{\chi}\}$ from ten independent mock catalogs of $N_{\rm det}=200$ detected events each, overlaid in grey. Dashed lines mark the injected true values. The diagonal panels show the marginalized one-dimensional distributions obtained by combining all realizations.
    }
    \label{fig:figS4}
\end{figure}

Detection and mock parameter-estimation posteriors are generated by the public package \texttt{GWMockCat}~\cite{Farah:2023vsc}, with the $\chi_{\rm eff}$ based spin modeling extension introduced in Ref.~\cite{Ng:2018neg}. \texttt{GWMockCat} draws true source parameters from a specified population model, applies a single-detector observed SNR threshold to determine detection, and produces mock PE posteriors with realistic scatter in detector-frame chirp mass, symmetric mass ratio, effective spin, sky-angle factor, and signal-to-noise ratio. The spin sector is parameterized through $\chi_{\rm eff}$ under the assumption that only the primary BH carries angular momentum, so that $\chi_{\rm eff} = \chi_{1}/(1+q)$. With $q=1$ fixed by our choice of injected population, the symmetric mass-ratio is bypassed entirely, and $\chi_1 = 2\chi_{\rm eff}$. We use a realistic noise power spectral density from the LVK
injection campaign~\cite{Essick:2025zed} as input to \texttt{GWMockCat}, and adopt a detection threshold of $\rho_{\rm obs} \geq 12$, obtaining $N_{\rm det}=200$ detected events per catalog. Fig.~\ref{fig:figS3} shows the recovered posterior on $\{m_1, \chi_1, z, \Theta\}$ for a representative detected event with $\rho_{\rm obs}=12.7$. The injected true values lie within the 68\% credible region for all four parameters, and the posterior exhibits the standard distance–inclination degeneracy visible as a strong $z-\Theta$ correlation.

The selection function is estimated from a single set of $7.5\times 10^8$ mock injections. Injections are generated from the same mass and redshift models with $\alpha=2.3, m_{\rm min}=3\, M_{\odot}, m_{\rm max}=200\, M_{\odot}$, and the same redshift slope. The primary spin magnitude is drawn from a truncated normal $\mathcal{N}_{[0,1]}(0,0.5)$,
chosen broad enough to cover the injected spin distribution across the full range relevant for evaluating $\xi(\Lambda)$, with $\chi_2=0$ and aligned $\chi_1$ consistent with our spin model. Each injection is assigned an independent $T_{\rm age}$ drawn from $\pi(T_{\rm age})$, and detection is determined by $\rho_{\rm obs} \geq 12$.

Due to the limited number of events in a catalog, the inferred hyperparameters may depend on the specific catalog realization. We characterize the typical reconstruction capability using ten independently generated mock catalogs, each containing $N_{\rm det}=200$ events, drawn from the same injected population model, that are above the detection threshold. We apply the hierarchical inference pipeline to each catalog, with all ten catalogs sharing the same selection-function injection set described above. Fig.~\ref{fig:figS4} shows the recovered $68\%$ credible contours for the
superradiance parameters $\{\log_{10}(m_b/\mathrm{eV}), \sigma_{\mathrm{SR}}\}$
and the baseline natal-spin parameters $\{\mu_{\chi}, \sigma_{\chi}\}$ overlaid across all ten realizations. The injected boson mass,
$\log_{10}(m_b/\rm{eV})=-12.1$, is recovered within the $68\%$ credible region in every realization, demonstrating that our inference pipeline provides an unbiased recovery of the central physical hyperparameter at the catalog size available here. The environment-induced spin dispersion $\sigma_{\rm SR}$ is also recovered around its injected value, albeit with broader posteriors and larger realization-to-realization scatter than $m_b$. However, according to Fig.~\ref{fig:figS4} , the baseline spin parameters $\mu_{\chi}$ and $\sigma_{\chi}$ are only recovered with broader posteriors and exhibit a larger realization-dependent variation. These parameters are intrinsically more difficult to measure. The data primarily constrain the superradiance induced spin distribution, leaving the baseline spin parameters only weakly restored. Note that this realization-dependent behavior does not propagate to the inference of $m_b$, which remains tightly constrained around the injected value across all realizations. The mass-function and redshift hyperparameters, not shown, are likewise recovered tightly around their injected values. 

\begin{figure}[ht]
    \centering
    \includegraphics[width=0.49\linewidth]{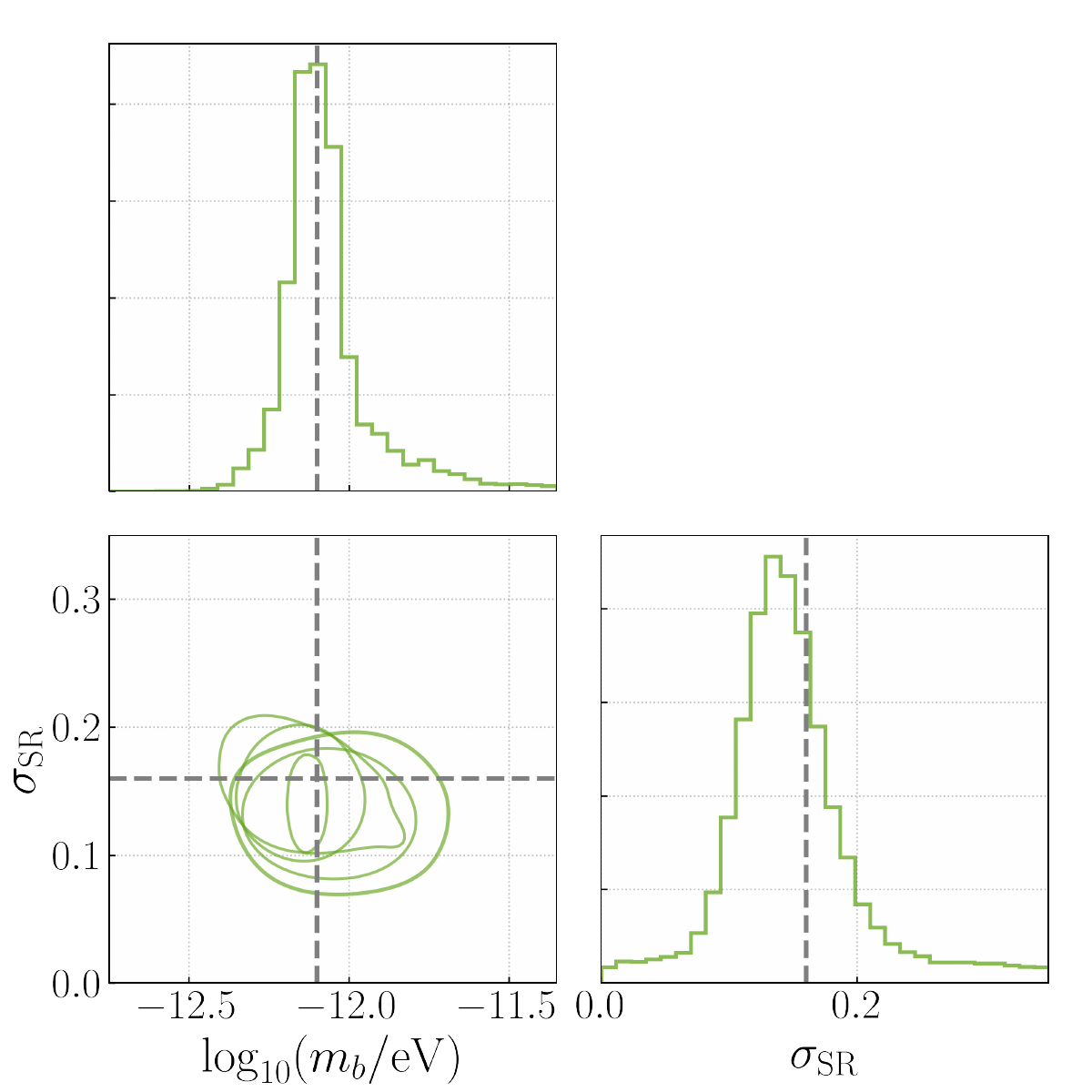}
    \includegraphics[width=0.49\linewidth]{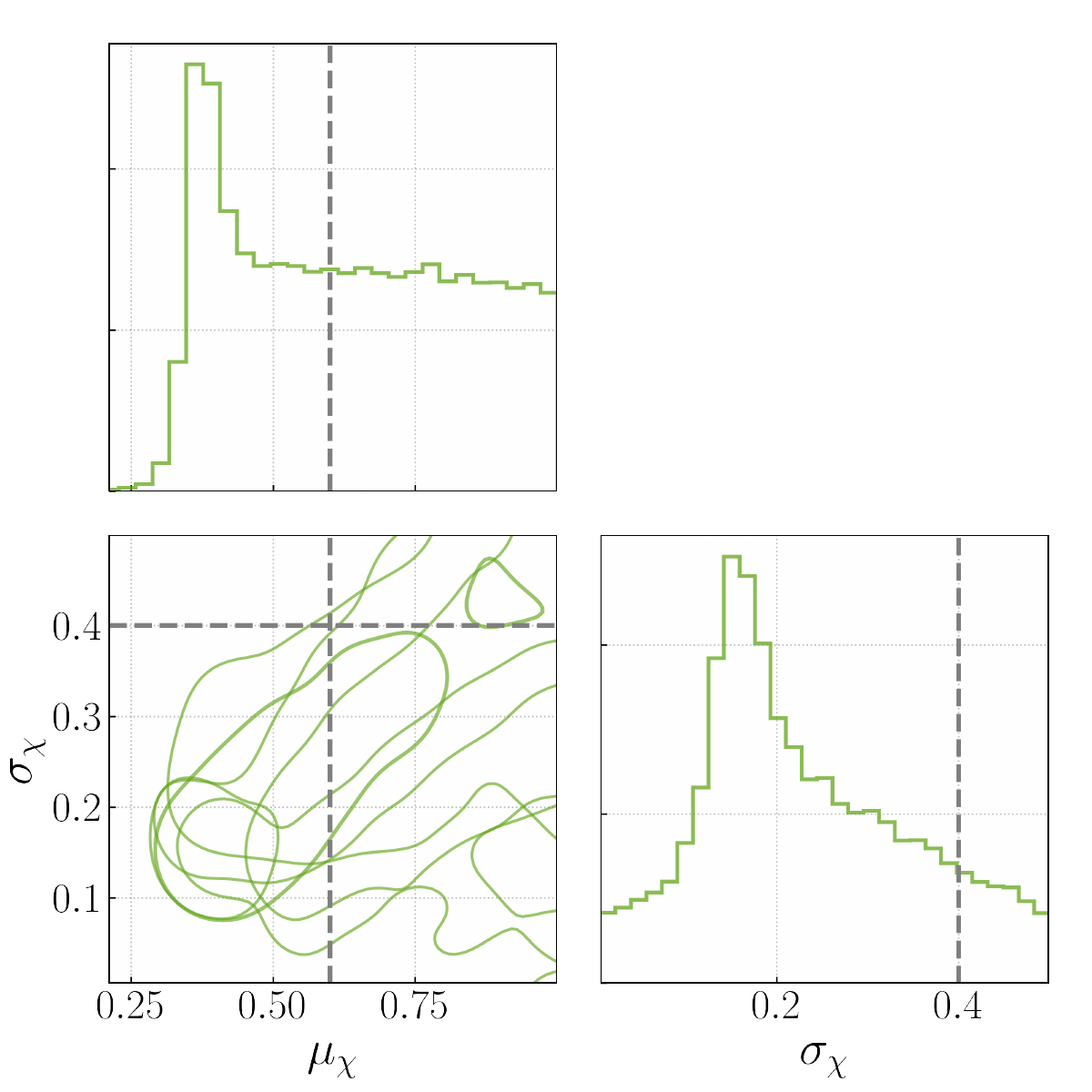}
    \caption{Reconstructed $68\%$ credible contours on the superradiance parameters $\{m_b, \sigma_{\mathrm{SR}}\}$ and the baseline natal-spin parameters $\{\mu_{\chi}, \sigma_{\chi}\}$ from five independent mock realizations in which the mass function is modified by $m_{\rm min} = 14.56\, M_{\odot}$ and $\alpha=4.8$, with all other injected hyperparameters unchanged. Dashed lines mark the injected true values. The diagonal panels show the marginalized one-dimensional distributions obtained by combining all realizations.}
    \label{fig:figS5}
\end{figure}

In principle, the boson mass should be approximately decoupled from the BH mass function, implying that valid Bayesian inference should yield unbiased estimates of the boson mass irrespective of the shape of the BBH mass function. To verify this, we repeat the mock data analysis with the mass function peak shifted from $\sim 10\, M_{\odot}$ to $\sim 20\, M_{\odot}$ by changing $m_{\rm min}$ from $4.56\, M_{\odot}$ to $14.56\, M_{\odot}$ and adjusting the power-law slope to $\alpha=4.8$, with $\delta_m$ held fixed (brown curve in Fig.~\ref{fig:figS2}). The injected boson mass remains to be $\log_{10}(m_b/\rm{eV})=-12.1$, and all other injected hyperparameters are unchanged. Fig.~\ref{fig:figS5} shows the recovered $68\%$ contours from five independent realizations of the shifted population, plotted in the same style as Fig.~\ref{fig:figS4}. The injected boson mass is recovered with precision comparable to the unshifted case despite the population mass peak now lying at $\sim 20\, M_{\odot}$, and $\sigma_{\rm SR}$ is likewise recovered around its injected value, while $\mu_{\chi}$ and $\sigma_{\chi}$ show the same weak recovery seen above. The boson-mass constraint reported in the main text is therefore not an artifact of any particular feature of the GWTC-3.0 to GWTC-5.0 mass function.

We restrict the mock data analysis to equal-mass binaries with $q=1$ and $\chi_2=0$, which focuses the test on the recovery of the $\chi_{\rm SR}(m_1)$ relation. Relaxing these assumptions to allow unequal masses and two spinning BHs is left to future work.


\section{Baseline spin hyperparameters}
\begin{figure}[h]
    \centering
    \includegraphics[width=0.49\linewidth]{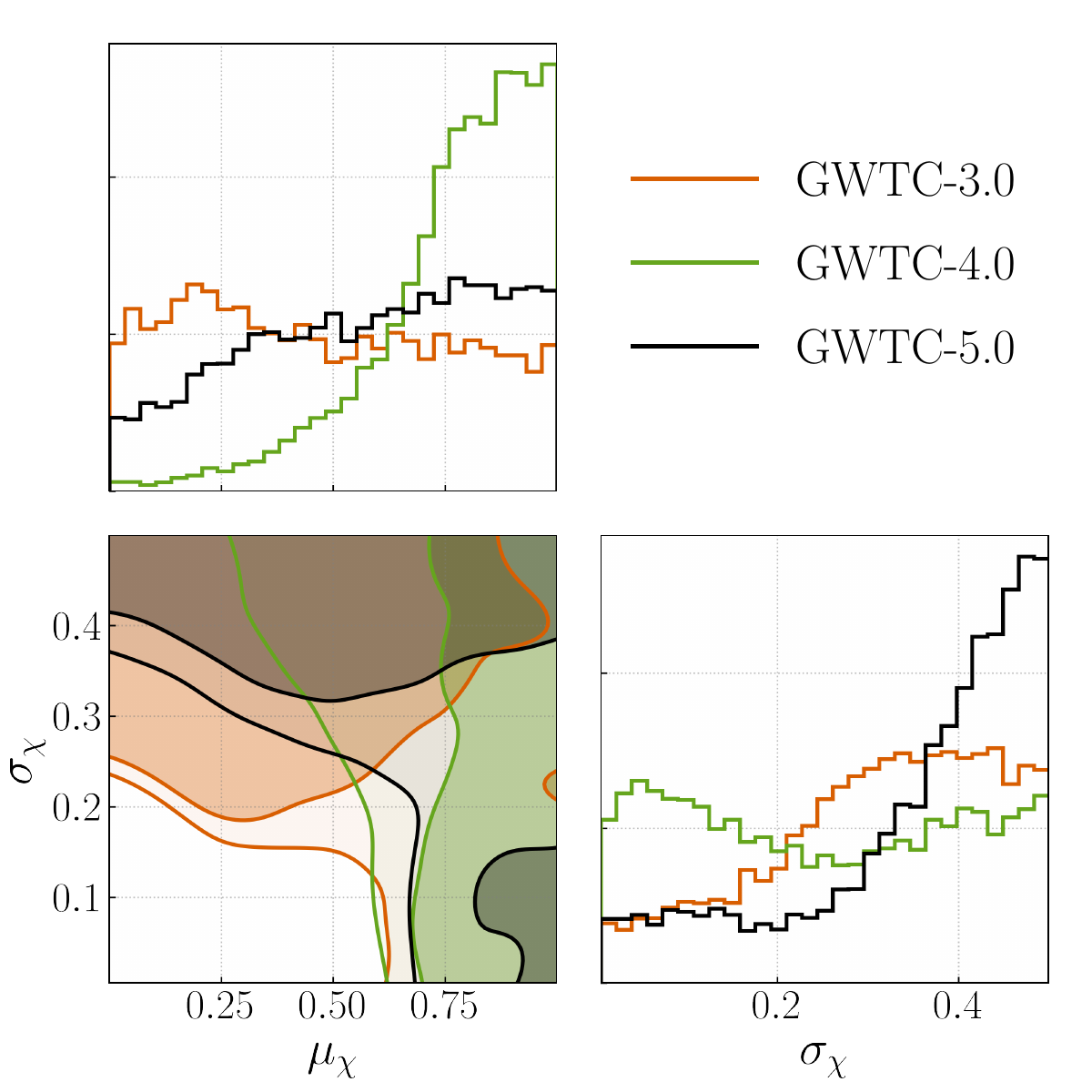}
    \includegraphics[width=0.49\linewidth]{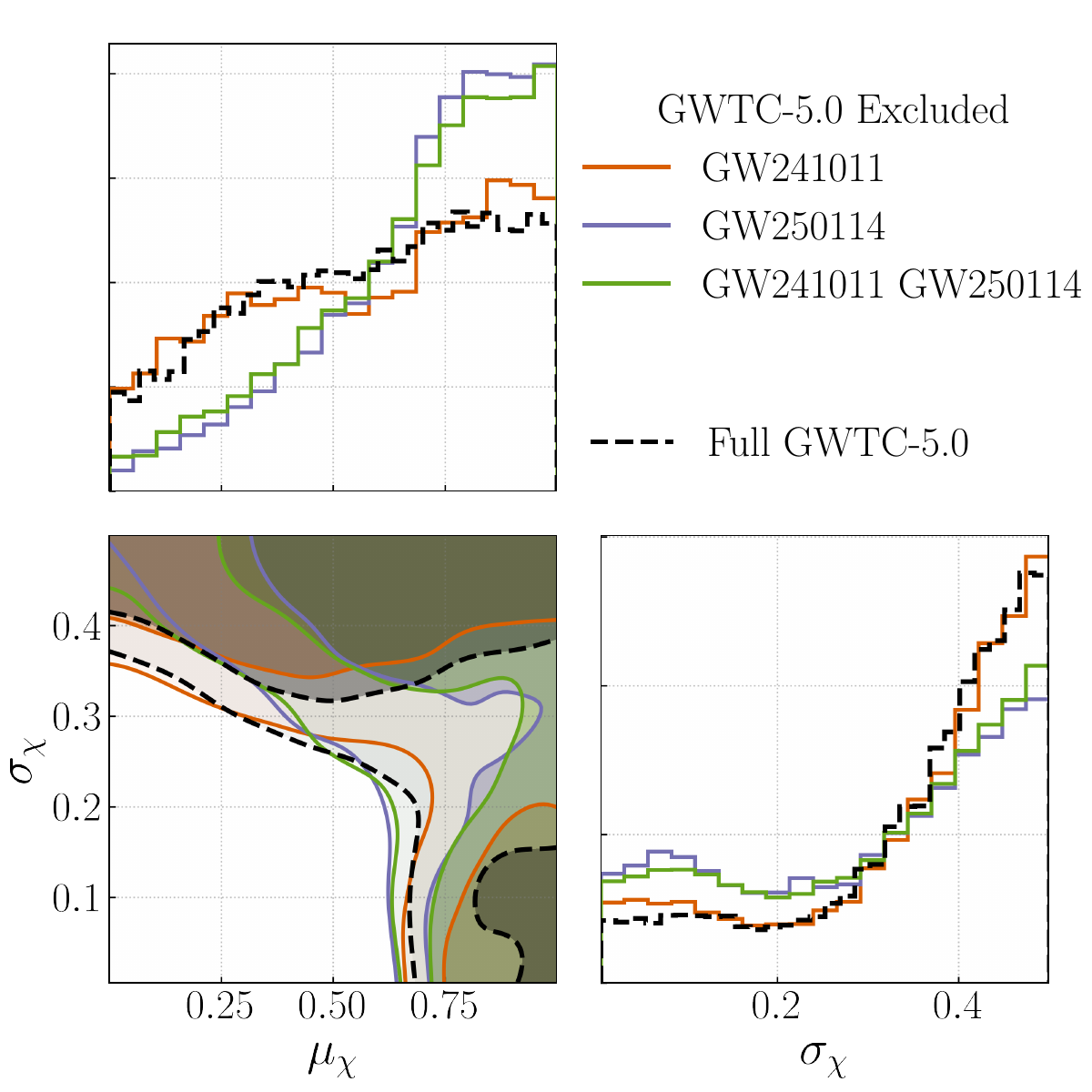}
    \caption{Posteriors on the baseline spin hyperparameters $\mu_\chi$ and $\sigma_\chi$ under the \textit{superradiance spin model}. Left:
    results for GWTC-3.0 to GWTC-5.0. Right: results from GWTC-5.0 after excluding GW241011, GW250114, and both events, compared with the full catalog. Contours enclose $68\%$ and $95\%$ credible regions.}
    \label{fig:figS6}
\end{figure}

For completeness, we report the posteriors on the baseline spin hyperparameters $\mu_\chi$ and $\sigma_\chi$ under the \textit{superradiance spin model}. The left panel of Fig.~\ref{fig:figS6} shows the results for GWTC-3.0 to GWTC-5.0, and the right panel compares the full GWTC-5.0 catalog with the exclusions discussed in the End Matter.

\section{Validation against the LVK GWTC-5.0 population inference}
\begin{figure}[h]
    \centering
    \includegraphics[width=1.0\linewidth]{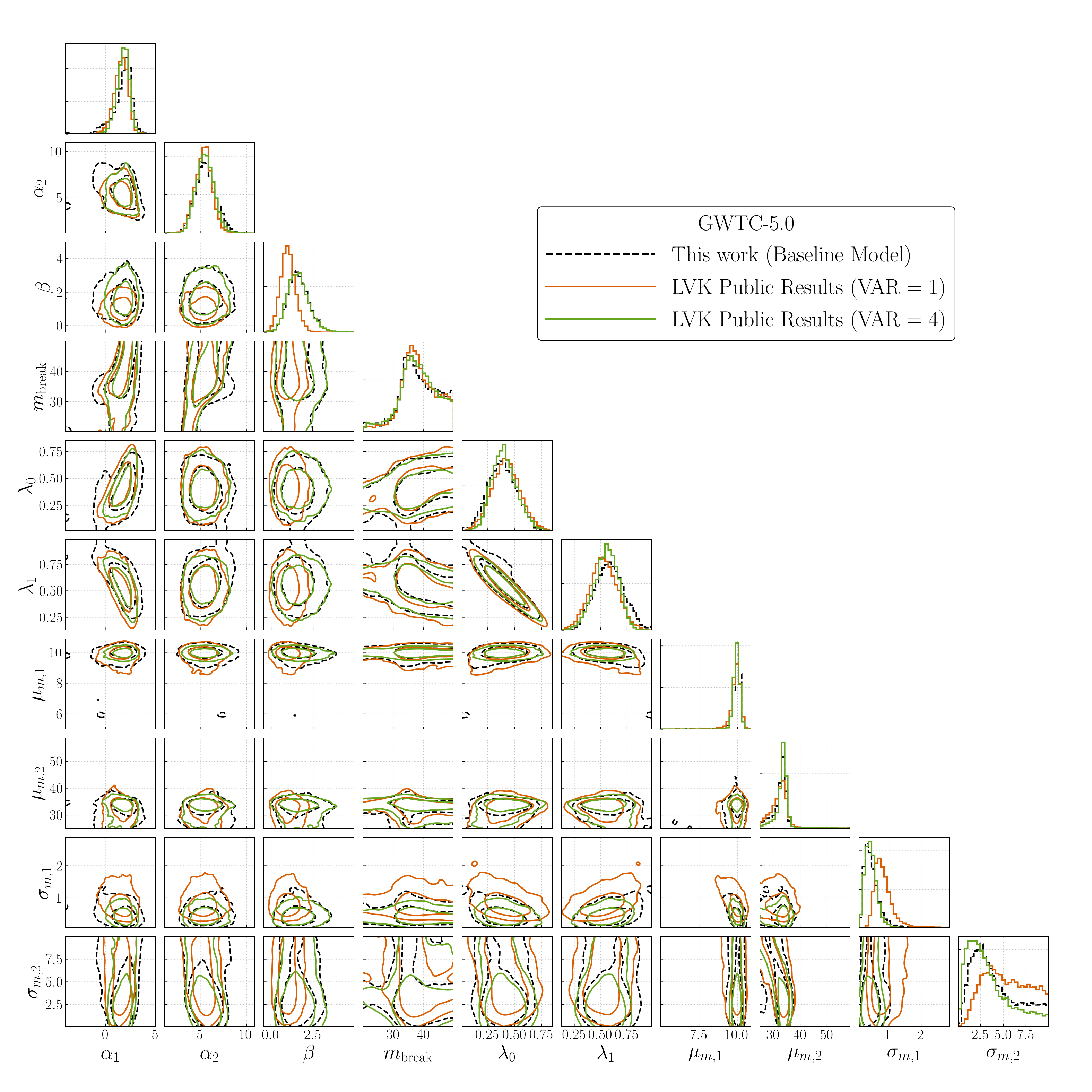}
    \caption{Posteriors on the hyperparameters of the \textit{Broken Power Law + 2 Peaks} mass model inferred from the GWTC-5.0 catalog. Black dashed contours show the results from our independent implementation; orange and green contours show the LVK public results~\cite{LIGOScientific:2026ctl} obtained under log-likelihood variance thresholds of $1$ and $4$, respectively. Contours show $68\%$ and $95\%$ credible regions.}
    \label{fig:figS7}
\end{figure}
Because our superradiance analysis performs the first fully joint hierarchical inference of the BBH mass function, redshift evolution, spin distribution, and superradiance parameters, we independently verify that our implementation of the LVK baseline population models reproduces the publicly released GWTC-5.0 results. This reproduction serves two purposes: it validates our pipeline against an established reference, and it provides log-evidences for the baseline model computed under the same numerical sampling settings as our superradiance runs, enabling the Bayes factor $\ln\mathcal{B}$ reported in the main text to be an apples-to-apples model comparison. We performed the hierarchical inference on the same GWTC-5.0 catalog and injection set with the baseline model described in the main text, using the \texttt{Dynesty} sampler at $n_{\rm live}=1000$ for improved convergence of both posteriors and log-evidence estimates.

The comparison also makes explicit the consequence of our choice of convergence criterion. As described in Sec.~\ref{sup:hierarchical}, we monitor the Monte Carlo accuracy of the hierarchical likelihood through the effective sample sizes of the per-event and selection-function sums, whereas the LVK analysis imposes a threshold on the variance of the log-likelihood estimator, $\sigma^2_{\ln\hat{\mathcal{L}}}$. Both approaches address the same underlying issue, and both are in common use, including within the same LVK analysis~\cite{LIGOScientific:2026ctl}, where an $N_{\rm eff}$-based criterion is adopted for some population models. A threshold on $\sigma^2_{\ln\hat{\mathcal{L}}}$ is a sufficient condition for unbiased inference~\cite{Essick:2022ojx, Talbot:2023pex}, but it is not a necessary one: for strongly parameterized population models a strict threshold can exclude regions of parameter space on numerical rather than physical grounds, while an $N_{\rm eff}$-based criterion is expected to give reliable results in the same setting~\cite{Heinzel:2025ogf}. Because the LVK analysis reports posteriors under two different variance thresholds, the two sets of public results bracket the effect of this choice.

\begin{figure}[h]
    \centering
    \includegraphics[width=0.6\linewidth]{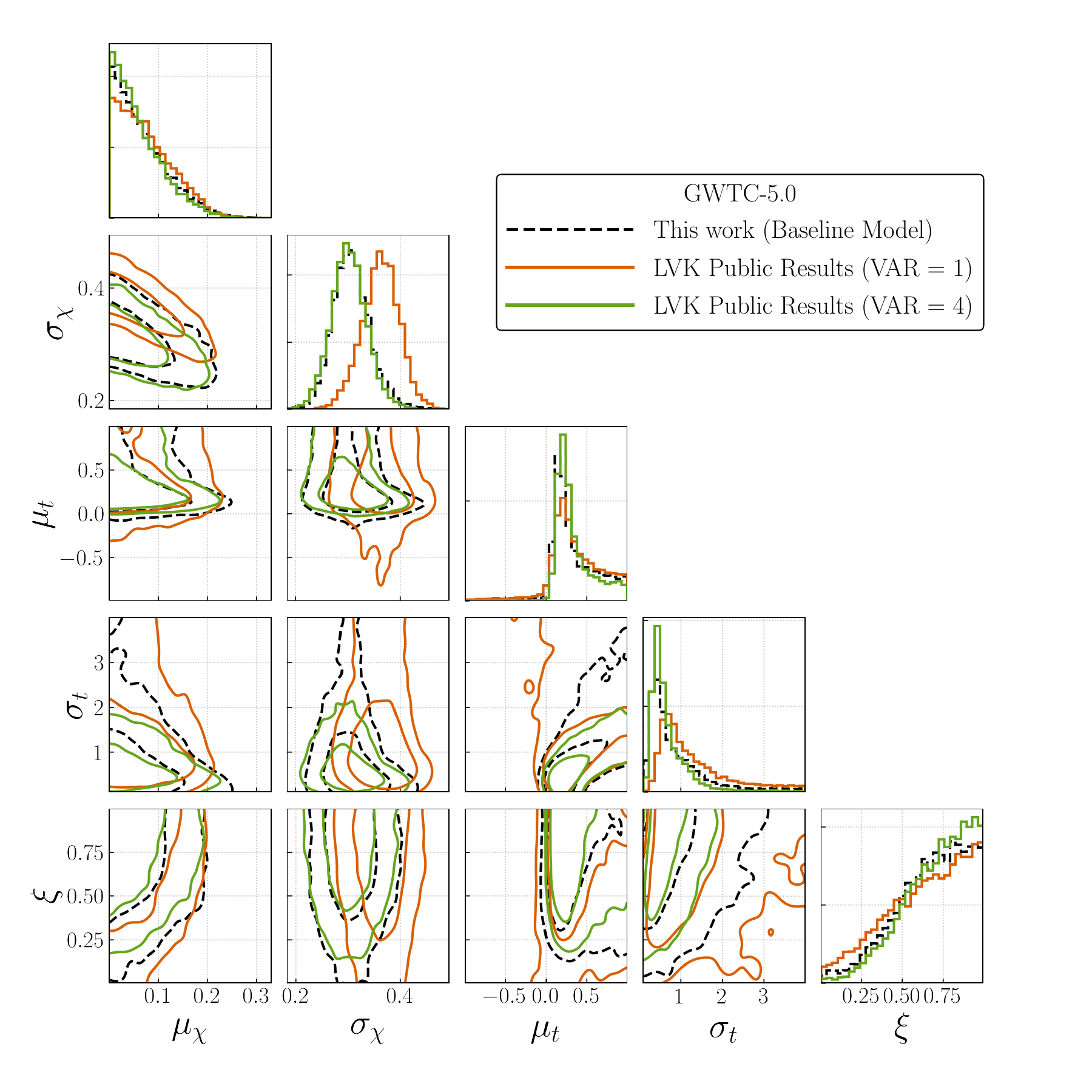}
    \caption{Posteriors on the five hyperparameters of the \textit{Gaussian Component Spins} model inferred from the GWTC-5.0 catalog: the spin-magnitude mean $\mu_\chi$ and standard deviation $\sigma_\chi$, the tilt-angle mean $\mu_t$ and standard deviation $\sigma_t$, and the mixing fraction $\xi$ between isotropic and preferentially aligned tilt components. Black dashed contours show the results from our independent implementation; orange and green contours show the LVK public results~\cite{LIGOScientific:2026ctl} obtained under log-likelihood variance thresholds of $1$ and $4$, respectively. Contours show $68\%$ and $95\%$ credible regions.}
    \label{fig:figS8}
\end{figure}

Figs.~\ref{fig:figS7} and \ref{fig:figS8} show the resulting posteriors on the mass-function and spin-distribution hyperparameters, overlaid with the LVK public results obtained under both variance thresholds. Our posteriors agree closely with the public results across all hyperparameters, and where the two LVK results differ, ours falls between them. This agreement establishes the reliability of the pipeline that we extend with the \textit{superradiance spin model} in the main text. We do not attempt to determine which convergence criterion is
preferable, and report this comparison so that the numerical choices underlying our inference are explicit. Mitigating the Monte Carlo uncertainty of hierarchical likelihood estimators remains an area of active development~\cite{Essick:2022ojx,Callister:2023tgi,Talbot:2023pex,Callister:2024qyq,Mancarella:2025uat,Hussain:2025llf,Hussain:2026pfm,Heinzel:2025ogf}, and methods that reduce the variance at its source rather than penalizing poorly sampled regions, such as the truncated Gaussian mixture representation of both single-event posteriors and selection injections discussed in the End Matter~\cite{Hussain:2025llf,Hussain:2026pfm}, offer a promising route to removing this ambiguity in future population analyses.

\end{document}